\documentclass{aa}  
\usepackage{textcomp}
\usepackage[utf8]{inputenc}
\usepackage{lscape}
\usepackage{lastpage}
\usepackage{amsmath,amssymb}
\usepackage{mathrsfs}	
\usepackage{multicol}
\usepackage{makecell}

\usepackage{graphicx}
\usepackage{txfonts}
\usepackage{hyperref}
\hypersetup{
    colorlinks   = true, 
    urlcolor     = cyan, 
    linkcolor    = blue, 
    citecolor   = blue 
}
\usepackage{placeins}
\usepackage{xcolor}
\usepackage{caption}

\graphicspath{plots}

{\newif\ifnotend
\notendtrue
\def\veclist{ABCDEFGHIJKLMNOPQRSTUVWXYZabcdefghijklmnopqrstuvwxyz.}
\def\top#1#2.{#1}
\def\tail#1#2.{#2.}
\loop\expandafter\xdef\csname bb\expandafter\top\veclist\endcsname%
{{\noexpand\bf\expandafter\top\veclist}}
\edef\veclist{\expandafter\tail\veclist}
\if\veclist.\notendfalse\fi\ifnotend\repeat}

\newcommand{\kpc}   {\,{\rm kpc}}
\newcommand{\pc}    {\,{\rm pc}}
\newcommand{\Mpc}   {\,{\rm Mpc}}
\newcommand{\Msun}  {\,M_{\odot}}
\newcommand{\Gyr}   {\,{\rm Gyr}}
\newcommand{\Myr}   {\,{\rm Myr}}
\newcommand{\yr}    {\,{\rm yr}}
\newcommand{\kms}   {\,{\rm km\,s^{-1}}}

\newcommand{\cm}    {\,{\rm cm}}
\newcommand{\K}     {\,{\rm K}}
\newcommand{\h}     {\,{\rm h}}

\newcommand{\rhogas} {\rho_{\rm gas}}

\newcommand{\rhodotstar} {\dot{\rho}_\star}
\newcommand{\epssfe}{\epsilon_{\rm SFE}}
\newcommand{\tauff} {\tau_{\rm FF}}
\newcommand{\tausf} {\tau_{\rm SF}}
\newcommand{\sigmasf}   {\sigma_{\rm SF}}

\newcommand{\Omegam}{\Omega_m}
\newcommand{\Ho}    {H_0}
\newcommand{\Mvir}  {M_{\rm vir}}

\newcommand{\Reff}  {R_\mathrm{e}}
\newcommand{\Mstar} {M_\star}
\newcommand{\Sigmao}{\Sigma_0}
\newcommand{\Sigmae}{\Sigma_{\rm e}}

\newcommand{\Mdm}   {M_{\rm dm}}

\newcommand{\Nclumps}   {N_{\rm clump}}
\newcommand{\Msttot}    {M_{\star, {\rm tot}}}
\newcommand{\Mcltot}    {M_{{\rm clump, tot}}}
\newcommand{\fcl}       {f_{\rm clump}}

\newcommand{\Sigmamod} {\Sigma_\star}
\newcommand{\Sigmaobsi} {\Sigma_{{\rm obs}, i}}

\newcommand{\Nbins}     {N_{\rm bins}}
\newcommand{\SigmaBG}   {\Sigma_{\rm BG}}
\newcommand{\Ri}    {R_i}
\newcommand{\RG}    {R_G}
\newcommand{\VG}    {V_G}

\newcommand{\qsmr}  {q_{R_{\rm eff}-M_\star}}
\newcommand{\msmr}  {m_{R_{\rm eff}-M_\star}}

\newcommand{\alphal}{\alpha_{50\,{Myr}}}

\newcommand{\tage}  {t_{\rm age}}
\newcommand{\tcross}{t_{\rm cross}}

\newcommand{\tevap} {t_{\rm diss}}

\newcommand{\dd}    {{\rm d}}

\newcommand{\SFH}    {{\rm SFH}}
\newcommand{\Sersic}{Sérsic\,}

\defcitealias{Calura2024}{C24}

\begin{document}

\title{SIEGE IV: compact star clusters in cosmological simulations with high star formation efficiency and sub-parsec resolution}
\titlerunning{SIEGE IV: compact star clusters in cosmological simulations}

\author{
R. Pascale \inst{1} \thanks{\email{raffaele.pascale@inaf.it}} \and
F. Calura \inst{1} \and
E. Vesperini \inst{2} \and
J. Rosdahl \inst{3} \and 
C. Nipoti \inst{4} \and
E. Giunchi \inst{4} \and 
E. Lacchin \inst{5,6,7} \and
A. Lupi\inst{8,1} \and
M. Messa \inst{1} \and
M. Meneghetti \inst{1} \and
A. Ragagnin \inst{1} \and 
E. Vanzella \inst{1} \and 
A. Zanella \inst{5}
} 
\institute{
INAF - Osservatorio di Astrofisica e Scienza dello Spazio di Bologna, Via Gobetti 93/3, 40129 Bologna, Italy 
\and
Department of Astronomy, Indiana University, Bloomington, Swain West, 727 E. 3rd Street, IN 47405, USA
\and
Centre de Recherche Astrophysique de Lyon UMR5574, Univ Lyon, Univ Lyon1, Ens de Lyon, CNRS, F-69230 Saint-Genis-Laval, France
\and
Dipartimento di Fisica e Astronomia ’Augusto Righi’, Università di Bologna, via Piero Gobetti 93/2, 40129 Bologna, Italy
\and 
Physics and Astronomy Department Galileo Galilei, University of Padova, Vicolo dell’Osservatorio 3, I–35122, Padova, Italy
\and
INFN - Padova, Via Marzolo 8, I–35131 Padova, Italy
\and
Institut für Theoretische Astrophysik, ZAH, Universität Heidelberg, Albert-Ueberle-Straße 2, D-69120, Heidelberg, Germany
\and
Como Lake Center for Astrophysics (CLAP), DiSAT, Univeristà degli Studi dell'Insubria, via Valleggio 11, I-22100 Como, Italy
}
\authorrunning{Pascale et al.}
\date{Received ...; accepted ...}
 
\abstract
{The formation of compact high-redshift star-forming clumps, the physical processes driving their evolution and their potential connection to present-day Globular Clusters are key open questions in galaxy formation. In this work, we aim to shed light on these aspects using the SImulating the Environment where Globular clusters Emerged (SIEGE) project, a suite of cosmological zoom-in simulations with sub-parsec resolution specifically designed to investigate the physical conditions behind the origin of compact stellar systems in high-redshift environments. The simulation object of this study focuses on a dwarf galaxy with a virial mass of a few $10^9\Msun$ at $z=6.14$, where the spatial resolution reaches $0.3\pc\h^{-1}$. Individual stars are formed directly by sampling the initial mass function with a 100\% star formation efficiency, a setup designed to explore the impact of a high star formation efficiency under high-redshift conditions. The simulation reveals the emergence of numerous stellar clumps with sizes of 1–$3\pc$, stellar surface densities up to almost $10^4\Msun\pc^{-2}$, and masses predominantly spanning from $10^3\Msun$ to several $10^4\Msun$, with a few reaching $10^5\Msun$ and up to $10^6\Msun$. All clumps form during intense, short bursts of star formation lasting less than a $\Myr$, without noticeable signs of second peaks of star formation or accretion, often with negligible dark matter content (dark-to-stellar mass ratios below 1 within three times their effective radii). We measure a clear correlation between mass and size, and a clump mass function described by a power-law with a slope of -2. Star formation conditions in the simulation behave similarly to those of a feedback-free starburst scenario, where dense clumps form due to inefficient stellar feedback over small timescales. Notably, some clumps exhibit properties closely resembling those of present-day globular clusters, highlighting their potential evolutionary connection.}


\maketitle

\section{Introduction}
\label{sec:intro}

The formation of globular clusters (GCs) remains an open question, with no definitive explanation yet agreed upon within the community. GCs are dense and compact stellar systems with no residual gas, characterized by typical stellar surface densities of $10^4\Msun\pc^{-2}$ and typical sizes of a few parsecs \citep{Baumgardt2020,Baumgardt2021}. A considerable fraction of GCs presents several chemical anomalies and/or multi-modal metallicity distributions \citep{Gratton2004,Carretta2009} that are nowadays interpreted as a signature of multiple populations (\citealt{Gratton2012}, \citealt{Bastian2018}, \citealt{Milone2022} and references therein). Also, the general consensus is that they lack dark matter \citep[e.g.][]{Moore1996,Conroy2011}. To form stellar clusters with these high densities and compact sizes, it is reasonable to speculate that specific conditions must be met. For instance, it is necessary to accumulate and convert large amounts of gas into stars efficiently within a short timescale and a confined volume, which requires environments of exceptionally high pressure and density \citep[pressure $P/k>10^7\K\cm^{-3}$ and density $n>10^4-10^5\cm^{-3}$;][]{Elmegreen1997,Kruijssen2014,Kruijssen2015,Renaud2015}. However, despite understanding these necessary conditions, many other aspects of GC formation remain unresolved. Key uncertainties include the mechanisms driving the formation of multiple stellar populations \citep{Bastian2018}, and whether GCs initially formed within dark matter halos \citep{Peebles1984,Boley2009,Trenti2015,Kimm2016,Renaud2018} that were later stripped by tidal interactions \citep{Mackey2005}, or if they formed in environments devoid of dark matter \citep{Naoz2014,Phipps2020}.

The majority of observed GCs is relatively old, with ages exceeding $10\Gyr$ \citep{Krauss2003,Forbes2010}, indicating that they must have formed in a young Universe, prior to reionization, where gas-rich and dense conditions were prevailing \citep{Tacconi2013,Wisnioski2015,Dessauges2019}. In such context, high-redshift observations are therefore essential, providing critical insights into the formation and evolution processes behind these objects. Indeed, data from the Hubble Space Telescope (HST) and, more recently, the James Webb Space Telescope (JWST) have enabled the discovery of numerous compact and massive sources in lensed fields at very high redshifts \citep{Vanzella2017,Vanzella2019,Calura2021,Mestric2022,Mowla2022,Vanzella2023,Claeyssens2023,Adamo2024,Messa2024a}. These sources exhibit extreme properties, with stellar densities up to $10^5\Msun\pc^{-2}$, or even $10^6\Msun\pc^{-2}$ confined in small physical sizes ranging from 2-$3\pc$ to a few tens of parsecs. While the exact nature of these dense systems remains uncertain, their similarities to present-day GCs suggest the possibility of an evolutionary link, implying that they could serve as progenitors of local GCs.

To enhance our understanding of these compact sources and their connection to GCs, and to complement the wealth of observational data available, it is crucial to develop a comprehensive theoretical framework that accurately captures the physical processes responsible for their formation but also considers the broader cosmological context,  essential to trace their evolution. In this respect, cosmological simulations are vital to bridge this gap in knowledge. However, to effectively characterize the small-scale dynamics occurring on parsec scales, they must achieve exceptionally high spatial and mass resolutions \citep{Kimm2016,Ma2020}, and must incorporate advanced feedback models \citep{Agertz2013,Dale2015,Ceverino2017,Weinberger2017,Pillepich2018,Hopkins2018,Marinacci2019}.

Historically, in the context of modeling small structures such as dwarf galaxies, most simulations have represented star clusters as single particles, operating at spatial resolutions of approximately $10\pc$ and mass resolutions of $10^3\Msun$. Such values may fail to adequately sample these small systems, leading to an inadequate modeling of the sub-parsec processes involved in star formation and the resulting feedback. To accurately capture their complexity, simulations must achieve sub-pc spatial resolutions, employ stellar particle masses of at most $10\Msun$ or, ideally, model individual stars that directly sample the initial mass function \citep[IMF; e.g.][]{Sormani2017} to get a proper stochastic representation of star formation \citep{Andersson2020,Calura2022,Deng2024}. However, fully resolving star clusters in early galaxies remains a challenge, primarily due to the dramatically increased computational costs associated with achieving such high resolutions across entire cosmic volumes.

Recent studies have advanced our understanding of the formation of GCs and their connection to high-redshift stellar clumps. As an example, by means of cosmological simulations, \cite{Ma2020}, \cite{Sameie2023} and \cite{Garcia2023} suggest that bound star clusters arise in high-pressure and high-density environments, highlighting the importance of the star formation model and how it influences cluster formation. Similarly, \cite{Calura2022} and \citet[][hereafter C24]{Calura2024} conducted zoom-in cosmological simulations with sub-pc resolution, examining the effects of various stellar feedback models on the formation of high-redshift clumps. The connection of GCs and dark matter has been addressed by \citet[][see also \citealt{Kimm2016}]{Gutcke2024} showing that multiple star formation bursts occur in low-mass dark matter halos and that dark matter can be efficiently lost by stripping, making these systems resemble GCs-systems.
However, despite these improvements, many limitations are still present. Most simulations stop at high redshift ($z>5-6$) due to the high computational cost (e.g. \citealt{Kimm2016}), or they fail to replicate the observed densities and/or sizes, or cover incorrect mass ranges, or a combination of thereof (e.g. \citealt{Ma2020}). The underlying reasons for these discrepancies are not entirely understood and may arise from various factors, such as resolution limitations, missing physical processes or processes inadequately implemented. Additionally, even when dealing with simulations that can successfully generate dense stellar aggregates with compact sizes and correct masses, they frequently struggle to accurately reproduce other dynamical or chemical signatures \citep{Kimm2016}.

In this work, we focus on the detailed analysis of a specific simulation within the framework of the SImulating the Environment where Globular clusters Emerged (SIEGE) project \citep{Calura2022}. This project encompasses a series of cosmological, hydrodynamical simulations with sub-parsec resolution, specifically designed to investigate the physical conditions that lead to the formation of dense stellar clusters. The high resolution of these simulations is required to model feedback processes at the scale of individual stars, which is crucial for accurately reproducing the complex dynamics that govern early star cluster formation.

The simulation analyzed in this work follows the evolution of a small cosmological volume centered around a massive dark matter halo, which is the progenitor of a dwarf galaxy, down to redshift $z=6.14$. At this redshift, the halo has a virial mass of approximately $4\times10^9\Msun$ (or, equivalently, $4\times10^{10}\Msun$ within three virial radii). The primary aim of this study is to investigate the structural properties of the stellar agglomerates formed within the volume and to assess how representative these clusters are of the dense stellar clumps observed in high-redshift lensed fields. To address this, we will analyze the scaling relations and structural properties of the clusters formed in the simulation and track their evolution as a function of redshift. As presented in \citetalias{Calura2024}, the feedback model implemented here features a high star formation efficiency, which strongly promotes the formation of compact and dense star clusters. This characteristic directly influences the compactness and density of the resulting systems (\citetalias{Calura2024}). 

This paper is structured as follows. In Section \ref{sec:setup}, we provide detailed information about the simulation parameters and setup. Section~\ref{sec:id} describes the algorithms employed to identify the stellar clusters, while Section~\ref{sec:struc} outlines the methods used to extract the structural parameters of these systems. In Section~\ref{sec:res}, we present our results, and in Section~\ref{sec:disc}, we discuss them in the context of existing literature. In Section~\ref{sec:concl} we draw our conclusions.

\begin{figure*}
    \centering
    \includegraphics[width=1\hsize]{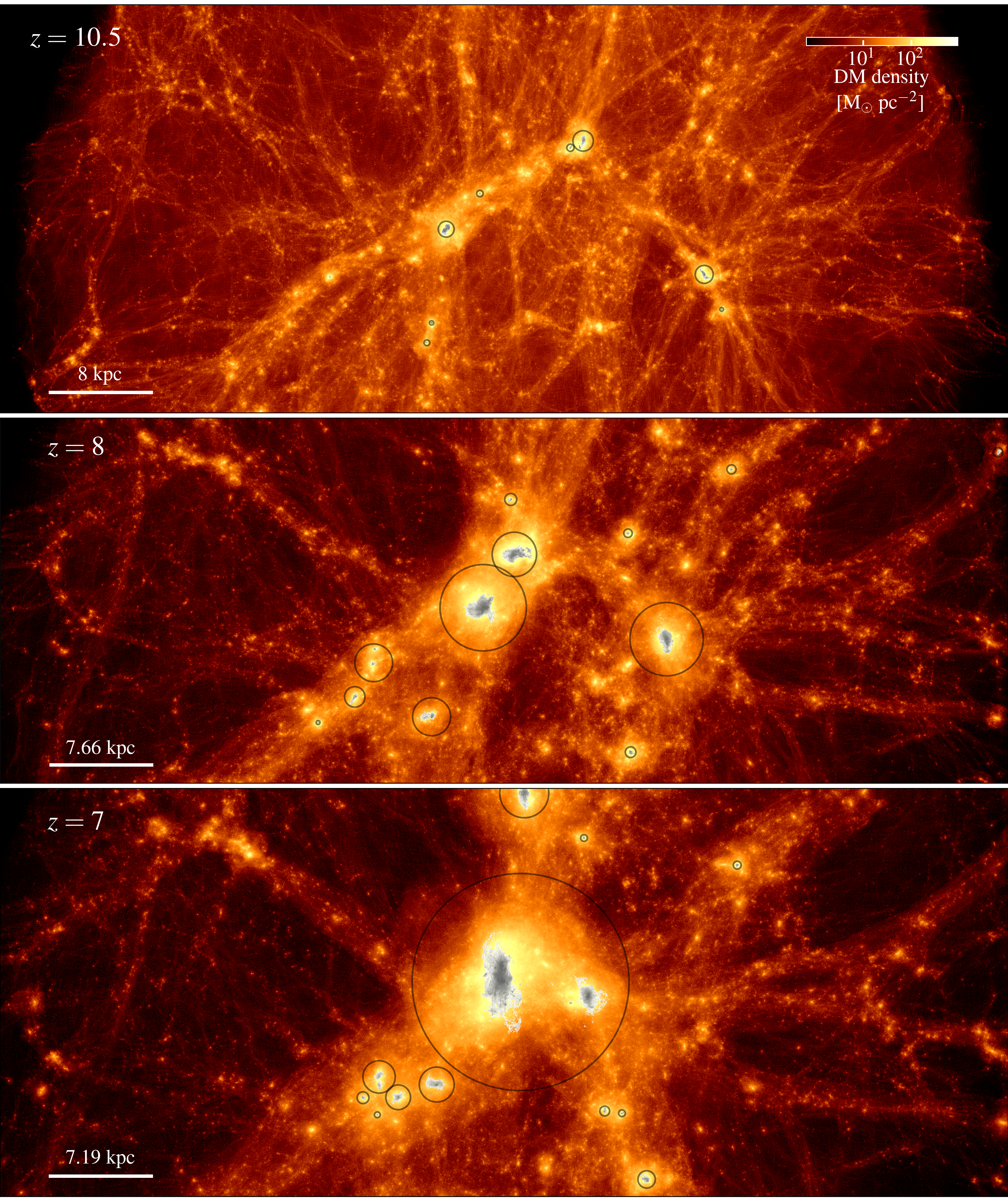}
    \caption{Evolution with redshift of the projected dark matter density within the high-resolution simulation domain. Top, middle and bottom panels show the projected dark matter density distribution in the central region of the simulation box at redshift $z=10.5,8$, and 7, respectively. In all panels, brighter colors indicate higher-density regions. The stellar projected density maps corresponding to the same regions are overlaid for comparison, allowing for a visual examination of the relationship between dark matter and stellar structures across different cosmic epochs. The stellar clumps formed have been highlighted with black circles.}
    \label{fig:fig1_dmstdens}
\end{figure*}

\section{Simulation set-up}
\label{sec:setup}

The simulation examined in this study was first introduced by \citetalias{Calura2024}, who, within the SIEGE framework, carried out and presented a series of cosmological simulations aimed at investigating how varying stellar feedback models influence the properties of the earliest star clusters formed at high redshift. The authors specifically examined the effects of different stellar winds models, varying star formation efficiencies, and distinct IMFs. Below, we summarize the main characteristics of the simulations presented by \citetalias{Calura2024}, with a focus on the one most relevant to our study. For further details regarding the simulation setup, additional information can be found in \cite{Calura2022}, \cite{Pascale2023}, and \citetalias{Calura2024}.

All simulations from \citetalias{Calura2024} were conducted using the adaptive mesh refinement (AMR) hydrodynamic code RAMSES \citep{Teyssier2002} and evolved down to $z=10.5$, with the exception of the one considered in this work, evolved down to redshift $z=6.14$. The simulations assume a flat $\Lambda$CDM cosmological model with a matter density of $\Omegam = 0.276$ and a Hubble constant of $\Ho = 70.3$ $\kms$ $\Mpc^{-1}$ \citep{Sharov2014,Omori2019}, a convention we will maintain throughout our analysis. The simulation box encompasses a $L=5\Mpc$ $h^{-1}$ comoving volume. The dark matter mass resolution is $165\Msun$ per particle in the zoom-in region, resulting in a total of approximately $2\times10^8$ dark matter particles. The use of the AMR scheme combined with a relatively small simulation box enables sub-parsec resolution, with a maximum refinement level of 21\footnote{The maximum level of refinement, $l$,  corresponds to the highest power of 0.5 applied to subdivide the edges of the simulation box when using AMR. In our case, the gas cell size at maximum refinement is $\Delta x = 0.5^l L$.}, corresponding to a spatial resolution of 0.3 $h^{-1}\pc$ at redshift 6.14.

As discussed in \cite{Calura2022}, all simulations of the SIEGE project were designed to reproduce the properties of the star forming complex observed in the lensed field strongly magnified by the galaxy cluster MACS J0416.1–2403 \citep[the {\em Cosmic Archipelago}; ][]{Vanzella2019,Calura2021,Messa2024b,Vanzella2024} at $z=6.14$. The complex comprises of several nucleated ($<10\pc$) stellar agglomerates, with masses from a few $10^5\Msun$ up to $5\times10^6\Msun$, with a total mass of no more than few $10^7\Msun$. On the basis of existing stellar-to-halo mass relations for low-mass halos \citep[][see also Fig. 2 of \citealt{Pascale2022}]{Ma2019}, a stellar system of this mass is, thus, expected to be hosted by a dark matter halo with a virial mass of a few times $10^9\Msun$, as in the simulation.

Fig.~\ref{fig:fig1_dmstdens} shows a representation of the evolution with redshift of the cosmic volume captured in the simulation. The top, middle and bottom panels illustrate, respectively, the central part of the simulation box at redshifts 10.5, 8 and 7. Throughout the panels, the filamentary structure of dark matter is clearly discernible. As the dark matter collapses, it forms increasingly massive knots that are likely to become sites of star formation. Each panel also includes the projected stellar density distribution of the relevant region for comparison with dark matter. The most massive halos visible at high redshifts begin to form stars at their centers, while being connected by dark matter filaments that gradually increase in size. The two massive halos visible in the top panel, linked by a dark matter bridge, move towards each other and nearly merge by redshift 8 (middle panel). By redshift 7, these halos have almost completely merged, and are about to accrete a third system forming the $\simeq10^9\Msun$ virial mass halo at $z=6.14$.

It is worth noting that despite different implementations of the feedback model, the large-scale properties of dark matter remain consistent across the various simulations of \citetalias{Calura2024}. Thus, although the dark matter density distributions in the figure are derived from the simulation analyzed in this study, they are representative of the general behavior across all simulations. Significant differences are, instead, observed in the small-scale properties of the stellar component (in terms of star formation histories, densities and  size of the formed stellar aggregates; \citetalias{Calura2024}).

The physical processes incorporated in the simulations include atomic radiative cooling from hydrogen, helium, and metals in photoionization equilibrium with a redshift-dependent ionizing ultraviolet (UV) background \citep{Haardt1996}, assuming reionisation starting at $z=8.5$. The simulations implement the formation of individual stars with the direct IMF sampling method \citep{Sormani2017}. Stars are formed at a rate
\begin{equation}\label{for:sf}
    \rhodotstar = \frac{\rhogas}{\tausf},
\end{equation}
with $\rhogas$ the gas density of a cell and $\tausf$ the timescale of star formation. The timescale of star formation is parametrized as $\tausf=\frac{\tauff}{\epssfe}$, with $\tauff\equiv\sqrt{\frac{3\pi}{32 G \rhogas}}$ the free fall time, $\epssfe$ the star formation efficiency per free fall time, and $G$ the gravitational constant. Gas cells eligible for star formation are those with a temperature below $2\times10^4\K$ and that, simultaneously, reach a density threshold that depends on redshift. The threshold is set requiring that no more than 90\% of gas in a cell is converted into stars, with this amount further limited to be, at most, $32\Msun$. Given a physical resolution of $0.3\pc$ at $z=6.14$, the condition translates into a density threshold of $10^4$ particles $\cm^{-3}$ (\citealt{Yaghoobi2022}; \citetalias{Calura2024}). At such densities and scales, gas typically available for star formation is enough for a few stars only, which motivates the need for individual star formation. Since stars are formed individually, the simulations incorporate feedback from individual stars in the form of stellar winds (SWs) and supernovae (SNe). Also, to prevent artificial radiative losses of the energy injected by the stars, a delayed cooling mechanism is employed \citep{Teyssier2013}, ensuring an avoidance of numerical overcooling and associated over-abundant star formation.

\begin{figure*}[h!]
    \centering
    \includegraphics[width=1\hsize]{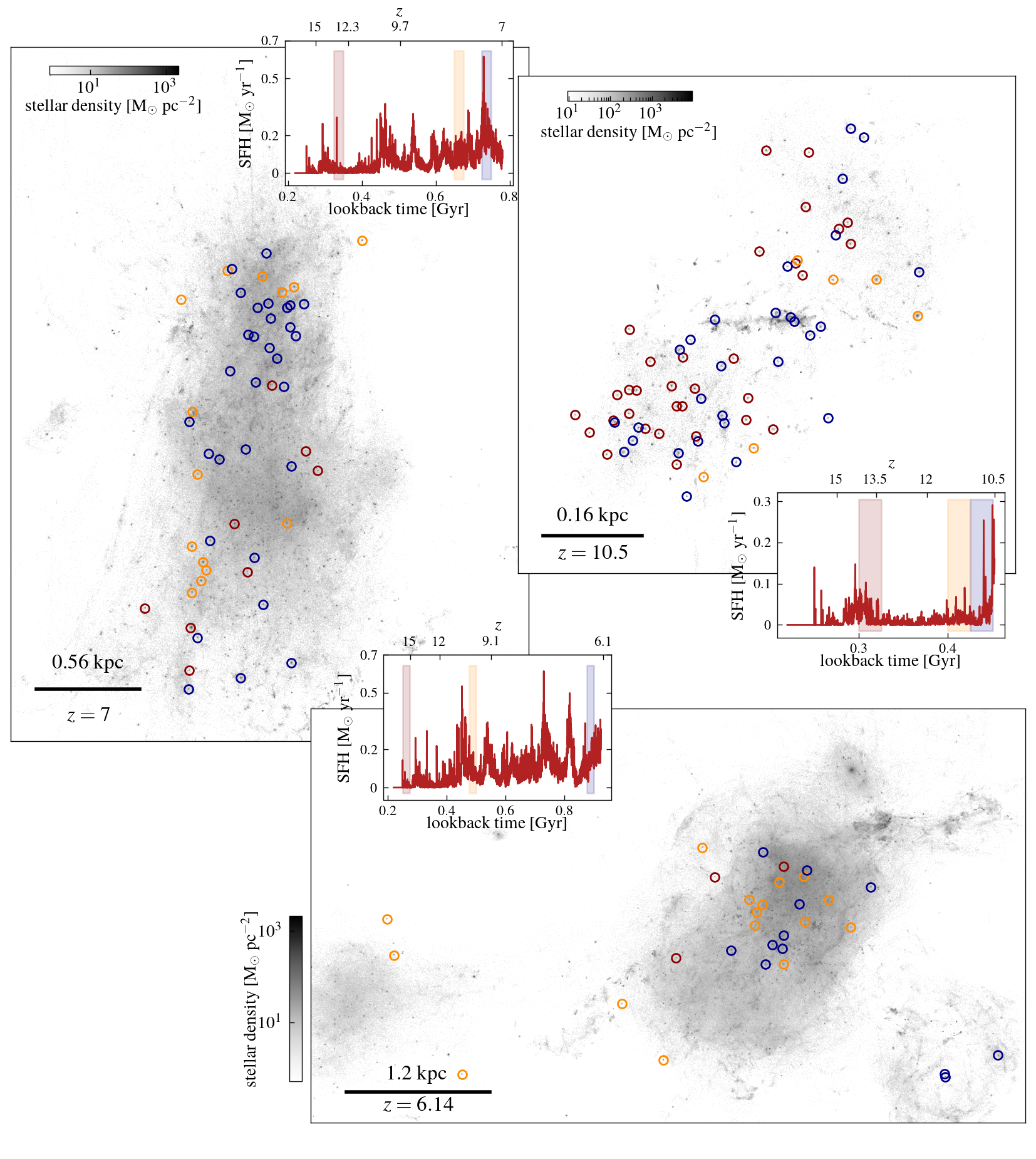}
    \caption{Examples of stellar density distributions of star forming regions taken at three different redshifts. The top right, top-left and bottom maps correspond to $z=10.5$, 7 and 6.14, respectively. The star forming systems shown in the top right and top left panels correspond to the stellar counterparts of the most massive halos visible in the top and bottom panels of Fig.~\ref{fig:fig1_dmstdens}. In each panel, the small insets display the star formation histories of the corresponding regions. All clumps identified within the time windows color-coded in the small insets are marked with circles and displayed with the same colors on the corresponding projected stellar density map.}
    \label{fig:fig2_stdens}
\end{figure*}

The simulations in \citetalias{Calura2024} are four and they differ in the implementation of specific aspects of the star formation and/or stellar winds model. Following the nomenclature established in \citetalias{Calura2024}, these simulations are (all sharing the same initial conditions):

{\em FC22}: The {\em FC22} simulation investigates the impact of stellar winds from massive stars implemented as an impulsive release of energy and mass into the interstellar medium (ISM) at the time of the formation of a stellar particle. The IMF is sampled according to a \cite{Kroupa2001}, with a star formation efficiency of $\epssfe=0.1$.

{\em WINDS}: Compared to the {\em FC22} simulation, the {\em WINDS} simulation modifies the stellar winds model by adopting a continuous injection of energy into the surrounding gas throughout the lifetime of the massive stars. The star formation efficiency is $\epssfe=0.1$ and the IMF is a Kroupa.

{\em THIMF}: In the {\em THIMF} simulation, the stellar winds model is consistent with that used in the {\em WINDS} simulation; however, the IMF employed is top-heavy, described by 
\begin{equation}
\phi(m) = \frac{\dd N}{\dd m} = \text{constant}.
\end{equation}
This choice is motivated by studies indicating a potential overabundance of massive stars in early stellar populations \citep{Marks2012,Charbonnel2014,Denissenkov2014,Cameron2024}.

{\em SFE1}: In the {\em SFE1} simulation, the IMF follows a Kroupa, and the stellar winds model is consistent with that used in the {\em WINDS} simulation. However, this simulation adopts a star formation efficiency of $\epssfe = 1$, in contrast to the $\epssfe = 0.1$ assumed in the other simulations (see Section~\ref{subsec:feedback}).

Our focus is on the SFE1 simulation because, according to \citetalias{Calura2024}, it consistently generates a significant number of realistic high-density, compact stellar clusters. While \citetalias{Calura2024} offers a comparative analysis of all four simulations, our analysis specifically examines SFE1 due to its apparent relevance in reproducing the characteristics of these stellar agglomerates. This is qualitatively shown in Fig.~\ref{fig:fig2_stdens} which presents a selection of star-forming regions from this simulation. Each panel focuses on a different redshift and such clusters are clearly evident as numerous small, high-density knots scattered around the relevant regions. All these knots have a size of a few parsecs, and can reach surface densities of several $10^3\Msun\pc^{-2}$. A more thorough look into Fig.~\ref{fig:fig2_stdens} is postponed to Section~\ref{sec:id}.

\section{Identification of stellar clumps}
\label{sec:id}

A key aspect of this work is the accurate and efficient identification of small stellar agglomerates across the large simulated volume and across redshift. To this purpose we developed a method specifically tailored to this simulation. We focus all the subsequent analysis on eight representative redshifts, namely $z=15.5$, 13, 10.5, 10, 9, 8, 7, and 6.14. For each of the considered redshifts, the identification method operates in two distinct steps. First, the simulation volume is divided into non-overlapping regions by identifying the major areas where stars have been formed. In the second step, the stellar particles belonging to each of these regions - tracing massive dark matter halos - are grouped into age bins which are independently scanned to search for stellar clumps. Both steps make extensive use of the Hierarchical Density-Based Spatial Clustering of Applications with Noise \citep{Campello2013} algorithm, as implemented in the software library HDBSCAN \citep{McInnes2017}.

For any given $n$-dimensional dataset, HDBSCAN constructs a minimum spanning tree (MST) from the distance-weighted graph of the data points, once an appropriate metric has been assumed (Euclidean in our case). An MST is a tree that connects all data points (nodes) with the shortest possible total distance (edge weights) without creating any loops. After constructing the MST, HDBSCAN varies the density threshold to build a hierarchy of clusters. This allows the algorithm to detect clusters of different densities, making it particularly effective for datasets with complex and noisy structures. HDBSCAN is an extension of the Density-Based Spatial Clustering of Applications with Noise \citep[DBSCAN;][]{Ester1996,Schubert2017}, and one of its advatages is that it does not require to specify a linking length ($\epsilon$), i.e. a characteristic distance used to evaluate whether elements of the dataset belong to the same cluster, but it evaluates stable clusters based on how long they persists as the density threshold changes. Persistent clusters are considered features of the dataset.

\begin{figure*}
    \centering
    \includegraphics[width=1\hsize]{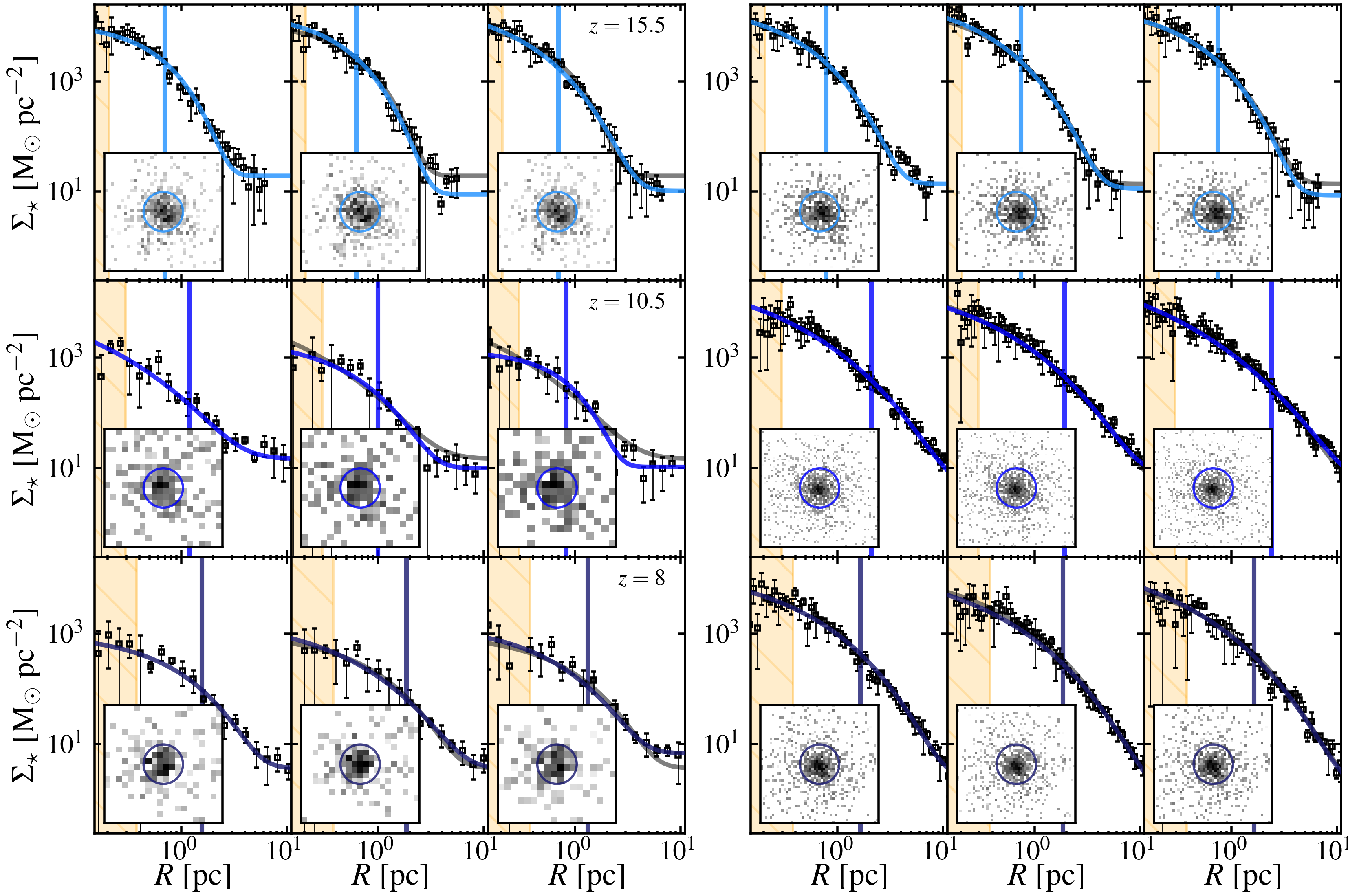}
    \caption{Clump projected density distributions. Top panels: Projected stellar density profiles (black squares with errobars) for two clumps selected at $z=15.5$. Each set of three panels shows the projected density along different lines of sight: along the $z$, $y$, and $x$ axes from left to right, respectively (the $xz$ plane is the plane of Fig.~\ref{fig:fig1_dmstdens}). The blue curves represent the best-fitting \Sersic + background models, while the vertical lines indicate the position of the effective radius $\Reff$. In the middle and right panels of each triplet we added, with a grey line, the \Sersic + background model shown in the left panel to highlight differences among projections. The curves and the binned profiles shown are those obtained at the end of the two step fitting procedure (see Section~\ref{sec:struc}). The small insets display the projected density map along the same direction as in the main panel, with the blue circle centered at center of the clump and having a radius equal to $\Reff$. The clump in the left panel has $(\Mstar, \Reff, n) = (1.25\times10^4\Msun, 0.67\pc, 1.24)$ , while the one in the right panel has $(\Mstar, \Reff, n) = (1.79\times10^4\Msun, 0.76\pc, 1.47)$. Middle panels: Same as the top panels, but for clumps at $z=10.5$. The clumps have $(\Mstar, \Reff, n) = (2.88\times10^3\Msun, 1.00\pc, 1.63)$ - left - and $(\Mstar, \Reff, n)$ = $(3.04\times10^4, 2.13\pc, 2.69)$ - right. Bottom panels: Same as the top and middle panels, but for clumps at $z=8$, with $(\Mstar, \Reff, n)=(3.03\times10^3\Msun, 1.57\pc, 1.62)$ and $(\Mstar, \Reff, n)=(1.83\times10^4\Msun, 1.71\pc, 2.27)$ - right. In each panel, the orange shaded area shows the simulation resolution ($\Delta x$) at that redshift, $\Delta x\equiv5 h^{-1}/[2^l(1+z)]$, with $l=21$ and $h=0.703$.}\label{fig:fig3_stars}
\end{figure*}

Delving into the details of our clump identification process, in the first step HDBSCAN is run on a randomly selected subsample of stellar particles. Using a subsample is essential to reduce computational costs, especially at lower redshifts, when the number of stellar particle's becomes extremely large ($>10^7$ at $z<8$). Here, we detect clusters directly operating on the MST setting a relatively large density threshold. A large density threshold in the MST allows to extract only few, well populated, clusters. Once a region is located, we determine its center as the median particles position and we define a characteristic radius as the largest Cartesian distance from the region’s center to any of its particles. For cases where the identified regions overlap or are concentric, we handle them as follows. Let $r$ and $R$ be the radii of two overlapping regions, with $r<R$, and $d$ be the distance between their centers; i) if $d<R-r$, the smaller region is eliminated, as it is entirely within the larger one; ii) if $R-r\le d<R$, the radius of the larger region is expanded to $R+r$ to include the smaller one which is, thus, removed; iii) if $R\le d<R+r$: the two regions are merged into one. The new center is calculated as a weighted average of the original centers (with weights proportional to the cubes of the original radii), and the new radius is set to the distance between the original centers.

As it will be discussed in details in the following sections, a key feature of the $\epssfe=1$ simulation is that star formation occurs in short, intense bursts. Each burst typically leads to the formation of a distinct stellar agglomerate. Thus, to enhance the effectiveness of our clump-finding algorithm, we leverage this characteristic by incorporating the ages of star particles into the clustering process. In the second step, for each identified region, we further group the stars into age bins of fixed width and run HDBSCAN separately on the particle positions within each age bin. 

HDBSCAN requires specifying one key parameter: the expected number of objects per cluster. In our implementation, this parameter is set to 500, which enables the identification of smaller-mass agglomerates. Given that stellar particles cannot be less massive than $0.5\Msun$, this setting allows for the identification of stellar agglomerates with a minimum mass of $250\Msun$. When applying HDBSCAN to the different age bins, each run typically identifies a large number of clusters. However, not all identified clusters correspond to genuine small, compact stellar agglomerates; some may represent stellar streams, isolated groups of particles, or background noise. Also, the same clump can, in principle, appear in different age bins if it experiences continuous or multimodal star formation. To accurately identify true agglomerates and avoid multiple detections, we employ the following refinement process. We first compute the center of mass for each identified cluster with an iterative scheme (shrinking sphere method; \citealt{Power2003}). Specifically we first get an initial estimate of the center of mass using only the stars classified as members by HDBSCAN within the relevant age bin. Subsequently, a refined center of mass is determined by including all particles located within a $5\pc$ radius of the previous center of mass estimate, regardless of their classification as members or age bin. We then assign to this mass center an average density computed using the 100 nearest particles, without any age distinction. In our framework, a cluster, thus, is identified in space by a single point representing its center of mass. Clusters are only retained if their average density exceeds a threshold of $25 \Msun \pc^{-3}$. The threshold of $25 \Msun \pc^{-3}$ is determined empirically: lower values tend to classify non-genuine agglomerates as clusters, as verified by visual inspection, while higher values tend to exclude progressively denser structures. As a final step, we perform a cross-match of the centers of mass within the star forming region of interest to identify possible multiple detections of the same clump. If a cross-match is detected, such as when the same stellar structure undergoes two or more star formation bursts, we eliminate one of the two detections.

The clustering algorithm is applied independently to all eight redshifts included in this study. As a result, clumps identified at redshift $z_i$ are not guaranteed to be identified in subsequent redshifts $z_{i+1}<z_i$. Also, we set the age bin width to $25\Myr$ but tested the clustering algorithm with a narrower bin width of $12.5\Myr$, obtaining convergent results regarding the number of identified agglomerates. After completing the entire identification procedure, we obtain a set of candidate clumps for each redshift. In the following Section, we will evaluate the structural properties of these clumps (see Section~\ref{sec:struc}) and conduct a second inspection based on these properties to identify and eliminate any potential remaining false detections.

Dividing the simulation box into star-forming regions and age bins offers several advantages. First, the algorithm runs approximately five times faster compared to executing it directly on all stellar particles in the simulation ($\simeq480$ CPU hours vs $\simeq80$ CPU hours at $z=6.14$). Second, it is significantly more efficient in terms of memory usage, allowing the algorithm to operate effectively without requiring high-memory computing clusters. Lastly, segmenting by age bins enhances cluster identification, leading to a higher number of distinct clusters. Without this age-based division, clusters can be challenging to differentiate, particularly when they appear as overdensities in areas of low particle density.

Fig.~\ref{fig:fig2_stdens} shows a selection of stellar projected density distributions of three star forming regions taken at different redshifts. The stellar map shown in the bottom panel is extracted from $z=6.14$ and it depicts the stellar system populating the central and most massive halo at this redshift. The top two panels correspond, instead, to $z=10.5$ (right) and $7$ (left), and illustrate the same stellar system as in the bottom panel, but during different evolutionary stages, before merging with other massive progenitors. Each star-forming region is populated by a multitude of small, dense knots, each extending over a scale of a few parsecs. Although not shown here for the sake of simplicity, these knots are present across the entire simulation, throughout every star-forming region, with maximum densities reaching nearly $10^4\Msun\pc^{-2}$. The abundance of these small and dense agglomerates  underscores the necessity for a highly effective clump-finding algorithm. The small panels attached to each projected density map display the star formation history of the corresponding region. The three vertical bands represent age bins, each spanning $25\Myr$. All clumps identified within these time windows are highlighted with circles and shown in different colors on the corresponding projected stellar density map. As anticipated, the star formation history is highly discontinuous, characterized by frequent bursts of star formation, as seen in the numerous peaks within these panels.

\section{Determination of stellar structural parameters}
\label{sec:struc}

The mass, size and any derived quantity (e.g. average surface density) of each identified clump are determined by fitting the clump's binned spatial density distribution to a well motivated model. The model surface density profile is parametrized by
\begin{equation}\label{for:sigma}
    \Sigmamod(R) = \Sigma(R) + \SigmaBG,
\end{equation}
where $\SigmaBG$ is a constant background added to a \cite{Sersic1968} surface density model
\begin{equation}\label{for:sersic}
\Sigma(R) = \Sigmao \exp \left[ -b(n) \left(\frac{R}{\Reff}\right)^{1/n}\right],
\end{equation}
with 
\begin{equation}
    \Sigmao = \frac{b^{2n}}{2\pi n \Gamma(2n)}\frac{\Mstar}{\Reff^2}
\end{equation}
and \citep{CiottiBertin1999}
\begin{equation}\label{for:bm}
    b(n) \simeq 2n - \frac{1}{3} + \frac{4}{405n} + \frac{46}{25515n^2}.
\end{equation}
In the previous equations, $\Gamma$ is the Gamma function, $\Mstar $ is the total stellar mass of the clump, $n$ is the Sersic index (a measure of the system's concentration), and $\Reff$ is the effective radius (i.e., the radius that contains half of the projected mass), such that
\begin{equation}
2\pi\int_0^{\Reff}\Sigma(R)RdR = \frac{\Mstar}{2}.
\end{equation}

\begin{figure*}[h!]
    \centering
    \includegraphics[width=1\hsize]{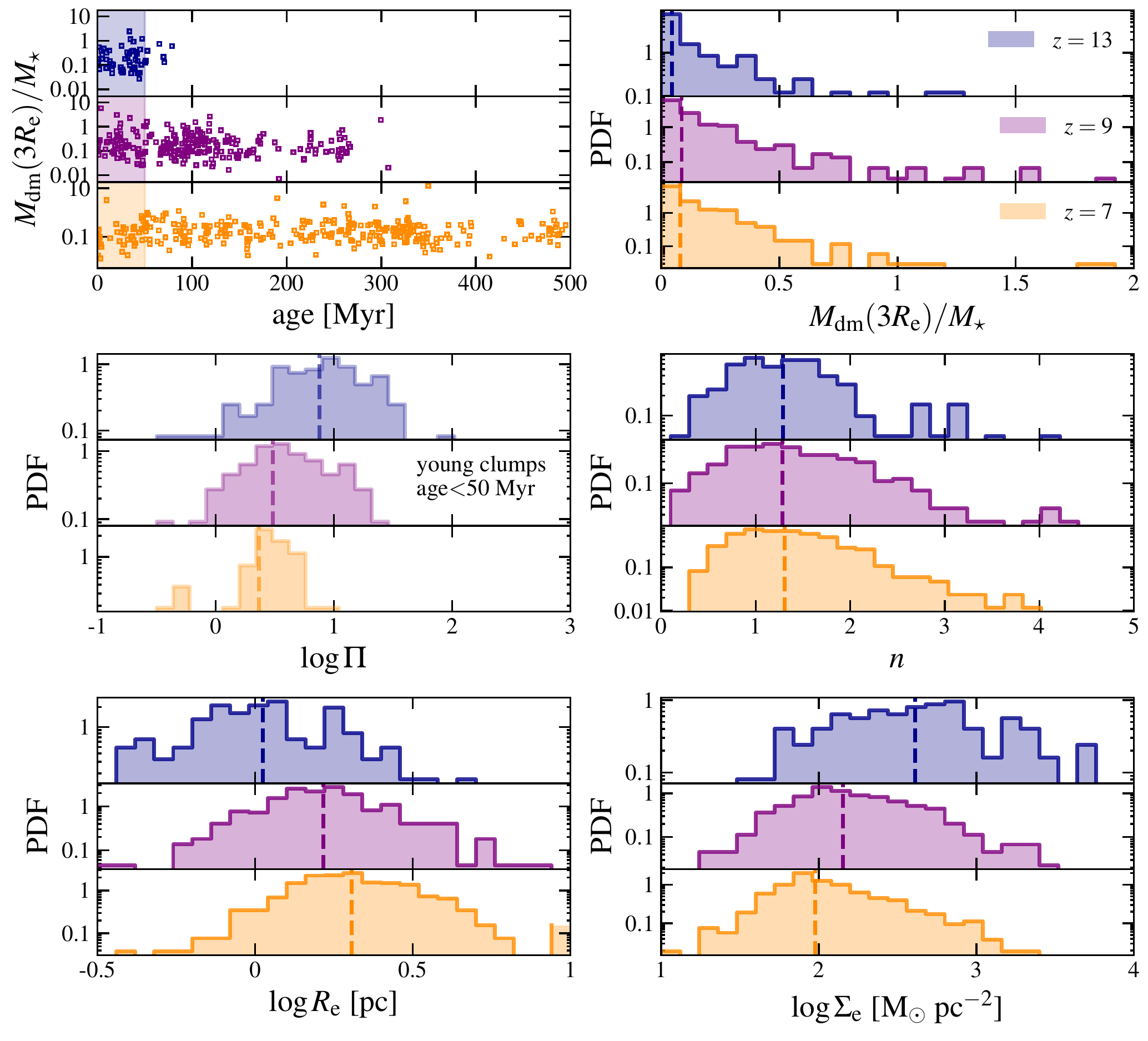}
    \caption{Clump parameters across redshifts. The triplets of panels illustrate: the relation between the dark-matter-to-stellar-mass ratio and clump ages (top left); the distributions of the dark matter mass within three effective radii relative to the total stellar mass, $\Mdm(3\Reff)/\Mstar$ (top right); the distributions of the boundedness parameter $\Pi$ for clumps younger than $50\Myr$ (middle left); the distributions of the effective radius $\Reff$ (middle right); the distributions of the \Sersic index $n$ (bottom left); and the distributions of the average surface density $\Sigmae\equiv\frac{\Mstar}{2\pi\Reff^2}$ (bottom right). Each panel within a triplet represents distributions at different redshifts, with $z=13$ shown in blue (top), $z=9$ in purple (middle), and $z=7$ in orange (bottom). The colored bands in the top left triplet of panels highlight clumps younger than $50\Myr$. These clumps are used to build the $\log\Pi$ distributions in the middle left triplet of panels. For clarity, only 3 of the 8 redshifts analyzed in this work are displayed. The vertical dashed line in all triplets except for the top left, show the median value of the corresponding distribution.}
    \label{fig:fig4_struct}
\end{figure*}

For each clump, we consider three independent projections, along the canonical $x$, $y$, and $z$ axes. For each projection, the fit is conducted as follows. First, we determine the number of radial bins ($\Nbins$) as the square root of the number of particles within a $10\pc$ radius from the clump center. The bins span the radial range $[0.1,10]\pc$, evenly spaced in logarithm. Using this binning scheme, we compute the clump's projected density profile which is fitted to get a first estimate of its effective radius. Using the derived $\Reff$, we adjust the radial bins to cover the range of $[0.1,10]\Reff$, we determine a new density profile and subsequently fit it again. The fit is performed by minimizing the figure of merit:
\begin{equation}\label{for:chi2}
\chi^2 = \sum_{i=1}^{\Nbins} \left( \frac{\Sigmaobsi - \Sigmamod(\Ri)}{\Delta\Sigmaobsi} \right)^2.
\end{equation}
Thus, for each projection of each clump, the total number of free parameter is four: $n$, $\Reff$, $\Mstar$ and $\SigmaBG$. The triplet $\{\Ri,\Sigmaobsi,\Delta\Sigmaobsi\}$ comprises, respectively, the average distance of the $i$-th bin, its surface density and uncertainty, and the sum in equation~(\ref{for:chi2}) extends over the total number of bins, with $i=1,...,\Nbins$. The minimization of equation~(\ref{for:chi2}) is performed using the Levenberg-Marquardt algorithm \citep{Levenberg1944,Marquardt1963}.

Following the determination of these structural parameters, we conduct a thorough and additional inspection of the clumps to further refine the sample by eliminating potential outliers. The exclusion criteria include clumps for which the Sérsic fit is unsuccessful in at least one of the three projections, those exhibiting half-mass radii estimates that deviate by more than a factor of two in one of the projections compared to the average, and any agglomerates misclassified as a large background region by visual inspection. These additional filtering steps refine the results and eliminates outliers. After applying these criteria, we obtain a sample of clumps, which is summarized in Table \ref{tab:tab1_params}.

After this process, each clump is assigned three triplets of total mass, effective radius, and Sérsic index, corresponding to fits from the three different projections. Since these quantities are inherently singular (e.g. a clump should have one mass, not three), we compute an average by taking the mean of the three fitted values, with error bars representing their dispersion\footnote{This dispersion serves as an indicative measure of the spread across projections, offering a practical first-order approximation. We stress that it does not represent a statistically robust measure of the spread, as it is based on only three values.}. The same method is applied to other derived quantities, such as the average surface density of the clump.

Fig.~\ref{fig:fig3_stars} presents a selection of clumps at three distinct redshifts: $z=15.5$, $13$, and $8$, with each row corresponding to one of these redshifts. For each redshift, we choose two clumps, a less massive one shown on the left and a more massive one on the right. Each clump is shown in three different projections, along with the results of the \Sersic fit. Also, in the small insets in each panel we show the corresponding projected stellar density distributions from which the profiles have been computed. Apart from a few cases, the clumps presented in the figure are broadly representative of the entire clump population and do not exhibit significant variations. The majority of clumps is very well-fitted by a \Sersic profile, indicating a consistent spatial structure (see also Section~\ref{subsec:struct}). Additionally, as it can be appreciated from the projected density maps, especially the most massive clumps tend to maintain an approximately spherical shape, without significant deviations in morphology.

To assign the additional characteristic property of age to the identified clumps we consider all particles within twice the average half-mass radius calculated across different projections, and determine the age as the mass-weighted average of the individual particle ages. This method ensures that the computed age accurately reflects the dominant stellar population within the clump. As we will demonstrate in detail, most clumps exhibit minimal age dispersion and have sharply peaked star formation histories (SFHs), rendering a very well defined concept of age.

\section{Results}
\label{sec:res}

Here, we present the key results of our work. We begin investigating whether the detected clumps are gravitationally bound and whether they form in dark matter halos (Section~\ref{subsec:dyn}). We analyze their structural properties (Section~\ref{subsec:struct}), study the evolution of various scaling relations with redshift, including the clump mass function (CMF, Section~\ref{subsec:cmf}) and the size-mass relation (Sections~\ref{subsec:smrelation}), and address the problem of clumps dissolution (Section~\ref{subsec:evap}).

\subsection{Dynamical state of stellar clumps}
\label{subsec:dyn}

Although the clustering algorithm is designed to detect dense and compact stellar systems, it does not inherently guarantee that these objects are gravitationally bound, as they may simply represent unbound stellar associations, especially right after their formation. To further investigate their dynamical state, we evaluate the parameter $\Pi$, an empirical measure of boundedness introduced by \cite{GielesZwart2011}, defined as
\begin{equation}\label{for:pi}
    \Pi\equiv\frac{\tage}{\tcross},
\end{equation}
where $\tcross$ is the crossing time and $\tage$ is the age of the clump. $\Pi$ reflects the relationship between a clump's dynamical time and its age, with values larger than 1 suggesting bound systems, and lower values indicating unbound or transient associations. The crossing time is estimated as
\begin{equation}
    \tcross = 10\sqrt{\frac{\Reff^3}{G\Mstar}}.
\end{equation}
The middle-left panels of Fig.~\ref{fig:fig4_struct} display the distribution of the $\Pi$ parameter at three representative redshifts: $z=13,9$ and 7, illustrating the evolution of the clumps' dynamical state across different epochs. In this case, we limited our analysis only to clumps younger than $50\Myr$ for two main reasons: i) in \cite{GielesZwart2011} the $\Pi$ parameter is calibrated only for objects below this age threshold; ii) clumps older than $50\Myr$ can reasonably be considered bound, as unbound systems would have dissolved on much shorter timescales. The values of $\Pi$ is almost always larger than 1 throughout the entire redshift range, indicating that the vast majority of these stellar clumps is not transient, unbound structures but rather form as gravitationally bound systems and maintain this state throughout their evolution.

\begin{table*}[t!]
\renewcommand{\arraystretch}{1.75}
\centering
\caption{Relevant properties of clumps across redshift.}\label{tab:tab1_params}
\begin{tabular}{ccccccccc}
\hline
\hline
$z$ & $15.5$ & $13$ & $10.5$ & $10$ & $9$ & $8$ & $7$ & $6.14$ \\
\hline
$\Nclumps$ & 4 & 104 & 207 & 320 & 369 & 408 & 436 & 358 \\
\hline
\makecell{Total stellar mass \\ $\Msttot$ [$10^6\Msun$]} & 0.07 & 1.81 & 6.58 & 12.56 & 19.08 & 28.49 & 46.83 & 72.27\\
\hline
\makecell{Total clump mass \\ $\Mcltot$ [$10^6\Msun$]} & 0.039 & 0.423 & 0.874 & 1.51 & 1.51 & 1.67 & 1.76 & 3.69\\
\hline
\makecell{Clump mass fraction \\ $\fcl$} & 0.57 & 0.23 & 0.13 & 0.12 & 0.08 & 0.06 & 0.04 & 0.05 \\
\hline
\makecell{CMF slope $\alpha$} & - & $-2.03^{+0.09}_{-0.09}$ & $-2.15^{+0.07}_{-0.07}$ & $-2.15^{+0.09}_{-0.09}$ & $-2.19^{+0.11}_{-0.11}$ & $-2.18^{+0.18}_{-0.18}$ & $-2.20^{+0.15}_{-0.15}$ & $-2.17^{+0.18}_{-0.18}$ \\
\hline
\makecell{young CMF slope $\alphal$} & - & $-2.07^{+0.19}_{-0.19}$ & $-2.02^{+0.08}_{-0.08}$ & $-2.02^{+0.15}_{-0.15}$ & $-2.15^{+0.15}_{-0.15}$ & $-1.98^{+0.23}_{-0.23}$ & $-1.60^{+0.18}_{-0.18}$ & $-1.73^{+0.22}_{-0.22}$ \\
\hline
\makecell{$\log\Reff-\log\Mstar$ relation  \\ $(\msmr,$ \\ $\qsmr)$} & \makecell{(0.32, \\ -1.48)} & \makecell{(0.44, \\ -1.59)} & \makecell{(0.46, \\ -1.55)} & \makecell{(0.30, \\ -0.93)} & \makecell{(0.36, \\ -1.07)} & \makecell{(0.53, \\ -1.57)} & \makecell{(0.38, \\ -1.02)} & \makecell{(0.39, \\ -1.08)} \\ 
\hline
\hline
\end{tabular}
\tablefoot{From top to bottom: the considered redshift $z$; the total number of clumps identified at the end of the procedure described in Section~\ref{sec:struc}, $\Nclumps$; the total stellar mass in the simulation volume $\Msttot$; the total stellar mass in clumps $\Mcltot$; the clump mass fraction $\fcl$; the slope $\alpha$ of the CMF ($\frac{\dd N}{\dd\Mstar}\propto\Mstar^{\alpha}$) of all clumps and the slope $\alphal$ of the CMF obtained considering only clumps younger than $50\Myr$; and the values of $\msmr$ and $\qsmr$, which result from the fits to the size-mass relation outlined in Section~\ref{subsec:smrelation}.}
\end{table*}

The issue of boundedness becomes more intriguing when considering dark matter. Whether GCs and massive high-redshift clumps actually form within dark matter halos remains a subject of debate. While the presence of dark matter may aid in forming bound GC-like objects \citep{Kimm2016}, the mechanism by which they would subsequently lose their dark matter to resemble present day GCs poses some challenges. Current theories suggest that, if formation within dark matter halos is the underlying  scenario, tidal stripping through interactions with other systems is the primary channel for dark matter loss \citep[e.g.;][]{Gutcke2024}.

Given that the $\Pi$ parameter in equation~(\ref{for:pi}) measures boundedness primarily for self-gravitating, single-component systems, it is essential to consider how and if dark matter affects our understanding of these clumps. To assess the contribution of dark matter, we estimated the ratio $\Mdm(3\Reff)/\Mstar$ for each clump and redshift, where $\Mdm(3\Reff)$ is the dark matter mass within three times the clumps' effective radius. The distribution of this ratio at the same three redshifts as for the distribution of $\Pi$ is shown in the top right panels in Fig.~\ref{fig:fig4_struct}. Across all redshifts and clumps, the ratio consistently remains low, rarely exceeding 0.5. The choice of $3\Reff$ as the characteristic scale for this measurement was motivated by the need to account for dark matter distribution while mitigating issues related to the discrete representation of dark matter particles in the simulation. This result underscores the minimal role of dark matter within the clumps, thereby justifying the use of equation~(\ref{for:pi}) to assess their bound status. The distributions in the top-right panels of Fig.~\ref{fig:fig4_struct} show that the clumps are not dark matter-dominated at the redshifts considered. However, they do not provide insight into whether the clumps initially formed within substantial dark matter halos and later lost dark matter due to processes such as stripping. To investigate this effect we show the relation between the ratio $\Mdm(3\Reff)/\Mstar$ and the clump ages in the top-left triplet of panels in Fig.~\ref{fig:fig4_struct}. If clumps were formed in dark matter-dominated halos and later experienced dark matter loss, we would expect a correlation where younger clumps exhibit higher ratios. However, as shown in the figure, no such correlation is evident at any redshift. This result strongly supports the conclusion that the clumps are currently in environments with minimal dark matter content and formed in regions that were already dark matter-poor.

This result is crucial as it demonstrates the simulation's ability to form relatively massive stellar systems, reaching masses as high as $10^5\Msun-10^6\Msun$ (see Section~\ref{subsec:struct}), without requiring substantial dark matter halos for formation and stability over time. This aspect will be investigated further in a future work (Pascale et al., in prep.).

\subsection{Structural properties of stellar clumps}
\label{subsec:struct}

The remaining triplets of panels in Fig.~\ref{fig:fig4_struct} offer distributions of key structural parameters of stellar clumps across redshifts, based on the \Sersic profile fitting of Section~\ref{fig:fig4_struct}. The panels display the distributions of the \Sersic index $n$ (middle right panel), effective radius $\Reff$ (bottom left panel), and average surface density (bottom right panel), defined as $\Sigmae\equiv\frac{\Mstar}{2\pi\Reff^2}$. The middle right panels reveal no significant evolution in the distribution of the \Sersic index $n$ with redshift. The distribution maintains a similar shape, with the index stable and ranging from $n=1.64_{-0.54}^{+0.64}$ at $z=13$ to $n=1.44_{-0.44}^{+0.59}$ at $z=7$, where the uncertainties represent the 16th and 84th percentiles of the distributions. These values span a wide range, indicating profiles with varying degrees of curvature. Small indices suggest cored profiles in the central regions and truncated in the outer parts, while large indices imply more centrally concentrated and shallower outer profiles.

While the distribution of the \Sersic index keeps constant across redshifts, the size distribution - bottom left panels - displays substantial evolution. At $z=13$, the distribution is peaked at smaller radii, with a median effective radius of $1\pc$, indicating that clumps formed at early times tend to be more compact. As redshift decreases the distribution shifts towards larger radii. The median value of the $\Reff$ distribution increases to $3\pc$, with some clumps reaching sizes of up to approximately $8\pc$.

\begin{figure*}[h!]
    \centering
    \includegraphics[width=1\hsize]{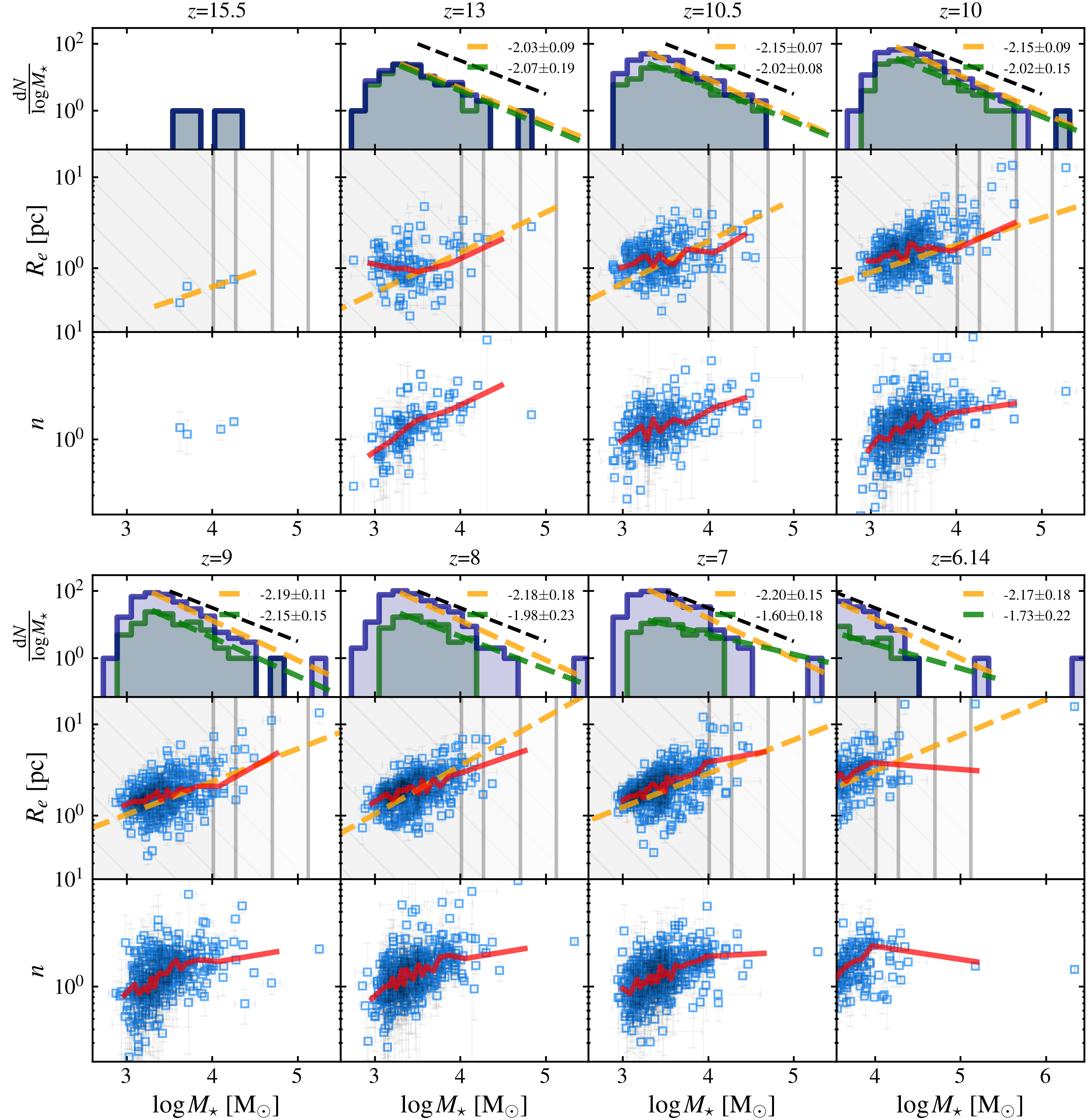}
    \caption{Scaling relations of clumps across redshifts. Top panels of each row: the CMF computed considering all (blue histogram) or only young ($\tage<50\Myr$, green histogram) clumps for the redshifts considered in this study. As a reference, the black dashed line indicates the $\frac{\dd N}{\dd\Mstar}\propto\Mstar^{-2}$ relation, corresponding to a scale free spectrum of structure formation. The dashed orange line shows, instead, the fit to the high mass end of the CMF ($\log\Mstar[\Msun]>3.25$), with the resulting slope indicated in the top right corner of each panel. Central panels of each row: size-mass relation as a function of redshift (blue points with error bars). In each panel, the orange dashed line corresponds to a linear fit to the data points. The grey vertical lines and bands represent the region of space where $\tevap\le13\Gyr$, with $\tevap$ as in equation~(\ref{for:evap}). Each line corresponds to a different assumption on $\RG$ in equation~(\ref{for:evap}), specifically, from left to right, $\RG=18\kpc$, $12\kpc$, $6\kpc$, $3\kpc$, while $\VG=220\kms$ to mimic the tidal field the clumps would experience if orbiting around a Milky Way like galaxy (see Section~\ref{subsec:evap} for detail). Bottom panels of each row: relation between the clumps stellar mass and \Sersic index. In the central and bottom panels, the red solid curve connects the median $\Reff$ and $n$ for each $\log\Mstar$ bin. Note that the panels  at $z=6.14$ cover a different mass range than the others to highlight the presence of a massive clump with $\Mstar\simeq2\times10^6\Msun.$}
    \label{fig:fig5_massfunction}
\end{figure*}

\begin{figure}
    \centering
    \includegraphics[width=1\hsize]{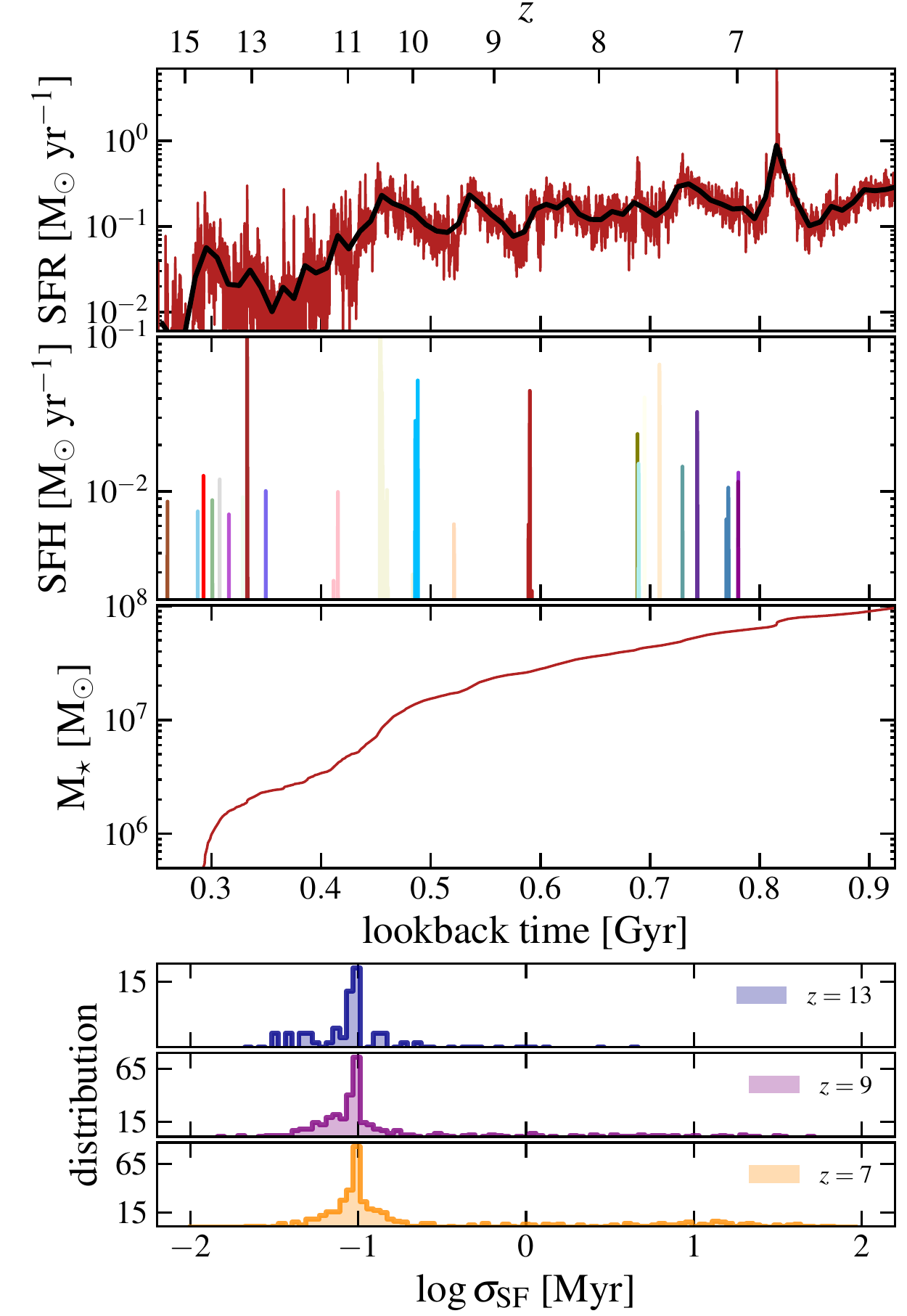}
    \caption{Top panel: SFH of the simulation at redshift 6.14, computed using two different binning schemes. The red line represents the SFH with bins of $0.2\Myr$, while the black line uses bins of $10\Myr$. 
    Middle panel; star formation histories of a selection of clumps at $z=7$. Each clump is color-coded differently, representing a range of variable masses and sizes. Bottom panel: Cumulative stellar mass as a function of time. Note that the cumulative mass represents the total stellar mass formed throughout the simulation and does not account for mass loss due to massive stars. Therefore, the effective mass at redshift 6.14 is smaller (see Table~\ref{tab:tab1_params}). The lower panels illustrate the distribution of $\log\sigmasf$, which represents the spread in the SFHs of all clumps (equation~\ref{for:sigsf}). This is presented for three distinct redshift values: $z=13$, $z=9$, and $z=7$.}
    \label{fig:fig6_overallSFH}
\end{figure}

The shift towards larger effective radii coincides with a decrease in surface density, as shown by the average surface density distributions shown in the bottom right panel of Fig.~\ref{fig:fig4_struct}. Indeed, while at high redshifts clumps exhibit very high average surface densities, with a median distribution of $460\Msun\pc^{-2}$ and few clumps approaching $6000\Msun\pc^{-2}$, as the redshift decreases, there is a notable shift towards lower surface densities. By $z=7$, the average surface density drops to $100\Msun\pc^{-2}$, with some values falling to as low as $30\Msun\pc^{-2}$.
This trend, along with the evolution of effective radii, suggests that stellar clumps appear larger over time, regardless of whether they form anew or evolve from pre-existing structures. This expansion may be attributed to two main factors: on one hand, the dynamic evolution, such as the hierarchical merging of smaller structures into larger objects; on the other hand, it could reflect a redshift dependence of gas conditions that promotes the formation of larger and less dense structures at lower redshifts. This aspect will be examined in greater detail in Section~\ref{subsec:evap}. Finally, we point out that additional processes not included in our simulations such as mass loss due to stellar evolution would cause the stellar systems to undergo an initial expansion. 

In the context of the study of the structural properties of stellar clumps, it is important to emphasize that the spatial resolution of our simulation (from $\simeq0.2\pc$ at $z=15.5$ to $\simeq0.5\pc$ at $z=6.14$) is close to the characteristic size of the smallest stellar clumps formed. While a formal convergence test would ideally require a higher-resolution run, such simulations are currently computationally prohibitive. We also note that lowering the resolution would not offer a meaningful test of robustness, as the reduced gravitational accuracy would artificially inflate clump sizes, preventing a reliable characterization of their structure. In this regard, the sizes we report should be interpreted as upper limits: resolving smaller spatial scales could potentially reveal even more compact and denser stellar clumps.

\subsection{Clump mass function}
\label{subsec:cmf}

A more quantitative analysis of the dependencies between structural parameters of clumps as a function of redshift is presented in  Fig.~\ref{fig:fig5_massfunction}. Here, the two sets of panels display, from top to bottom, the CMF, the size-mass relation, and the relationship between size and \Sersic index, while each column is a different redshift. We start focusing on the CMF shown at the top with blue histograms. In each panel we report with a black dashed line the $\frac{\dd N}{\dd \Mstar}\propto\Mstar^{-2}$ power-law relation, while the orange line indicates the best fit to the CMF in the relevant redshift bin, used to quantify changes in the CMF. We do not attempt to fit the CMF at $z=15.5$ since only four stellar clumps have formed at this redshift. The slope of the best-fit line is reported in the legend. It is generally expected that the CMF has a power-law behavior, with a typical slope of -2, reflecting the scale-free nature of gas fragmentation of structure formation \citep{Elmegreen1997,Guszejnov2018}. In our case the CMF keeps a slope consistently around -2 within the errorbars, ranging from -2.03 at redshift 13, down to redshift 7, where it is $\simeq-2.20$. We also show in Fig.~\ref{fig:fig5_massfunction} the CMF of young clumps, selected with an age criterion of $\tage < 50\Myr$, to make it comparable with observations, where clumps are identified through emission from young stars (e.g. UV emission). At the highest redshift ($z=13$), the CMF of young and all clumps almost coincide by definition, as all clumps are inherently young. As redshift decreases, the CMF of young clumps is also stable around a slope of $\simeq-2$, in agreement with observations of local \citep{Zhang1999,Hunter2003,deGrijs2003,Rodruck2023} or higher redshift young massive clumps \citep{Dessauges2018,Livermore2012}. The largest differences between the slopes of young and global CMFs are observed at the lowest redshifts, where the global CMF is increasingly dominated by older clumps, particularly at lower masses. Here ($z=7$ and $6.14$), the CMF of young clumps also deviates more than $1\sigma$ from -2, with a difference of 0.3-0.4. The corresponding slope values are provided in Table~\ref{tab:tab1_params}.

Unlike previous studies that observe CMF slope variations and link them to fluctuations in star formation activity \citep{Garcia2023}, we do not observe a significant trend in our simulation. In Fig.~\ref{fig:fig6_overallSFH} we show the SFH of the entire simulation box (top panel) alongside the cumulative mass formed as a function of time (bottom panel). The SFH is computed using bins of varying sizes, with the black curve representing larger bin widths to emphasize the overall mean trend rather than its burstyness, as highlighted by the red line. Up to approximately $z=11$, the SFH remains low, averaging a few times $10^{-2}\Msun\yr^{-1}$, a value modest compared to the higher star formation rates at lower redshifts. After $z=10.5$, the average star formation rate increases by almost a factor of 10, reaching typical values of $0.1\Msun\yr^{-1}$ and showing a fluctuating pattern with several bumps and sporadic short bursts of $\simeq0.5\Msun\yr^{-1}$. For instance, the increase in star formation activity at $t=0.75\Gyr$ ($z=7.3$) corresponds to the major merger of the two massive central stellar systems shown in the middle panel of Fig.~\ref{fig:fig1_dmstdens}. This merger event ultimately forms the large system depicted in the top left panel of Fig.~\ref{fig:fig2_stdens}.

At redshift 6.14, we excluded from the fit to the CMF the massive system with a stellar mass of $10^6\Msun$ visible in the bottom right panel of Fig.~\ref{fig:fig5_massfunction}. The formation of this massive system is also responsible for the peak in the SFR of $10\Msun\yr^{-1}$ visible in Fig.~\ref{fig:fig6_overallSFH} at redshift 6.8. This system exhibits singular characteristics, such as an unusually high mass and small size. Also, as highlighted in Section~\ref{subsec:dyn}, this clump forms without a significant contribution from dark matter. Given its peculiar properties and its striking structural and kinematic similarities to local GCs, a dedicated analysis of this system is addressed in a separate study (Pascale et al., in prep.).

\subsection{Size-mass relation vs redshifts}
\label{subsec:smrelation}

The central rows of panels to the top and bottom of Fig.~\ref{fig:fig5_massfunction} show the clumps size-mass relation as a function of redshift. A correlation between mass and size is evident, with smaller clumps populating the lower-mass region and larger clumps occupying the higher-mass region, although we consistently observe a large scatter in the relation. To quantify this dependence, we have included two curves in each panel: a linear fit in the $\log\Reff-\log\Mstar$ plane, shown with an orange dashed line, and a line connecting the median $\Reff$ computed in each mass bin, represented by a solid red curve. The linear fit has equation
\begin{equation}\label{for:sizemass}
    \log\Reff = \msmr\log\Mstar + \qsmr, 
\end{equation}
with $\msmr$ and $\qsmr$, the slope and normalization of the relation, respectively. As shown in Fig.~\ref{fig:fig4_struct}, at lower redshifts, the region containing larger sizes becomes more populated, suggesting the formation of increasingly massive structures. Consequently, also the clump size-mass relation exhibits a mild evolution with redshift: after a first increase of the slope in the redshift range $z=15.5-10.5$, where it evolves from 0.3 to 0.7, it stabilizes and it keeps approximately the constant value of 0.5 down to redshift 6.14. The values of $\msmr$ and $\qsmr$ are reported in Table~\ref{tab:tab1_params} for each redshift bin considered. As already noted in Section~\ref{subsec:struct}, the spatial resolution of the simulation is close to the characteristic size of the smallest clumps identified. As a result, being able to resolve even smaller spatial scales may potentially lead to a steepening of the relation, increasing the population of low-mass, small-sized clumps.

Although Fig.~\ref{fig:fig4_struct} (top right panel) indicated that the distribution of the \Sersic stays relatively constant across redshifts, the bottom rows of panels in Fig.~\ref{fig:fig5_massfunction} highlight a weak correlation, with a significant amount of scatter, between the \Sersic index and $\Reff$. In this case we did not provide any fit, but, to highlight this dependence, we show the median \Sersic index per mass bin. Smaller clumps tend to have a median \Sersic index close to 1, which suggests an exponential profile, while larger and more massive clumps display a higher median \Sersic index, with also less scatter. 

\begin{figure*}
    \centering
    \includegraphics[width=1\hsize]{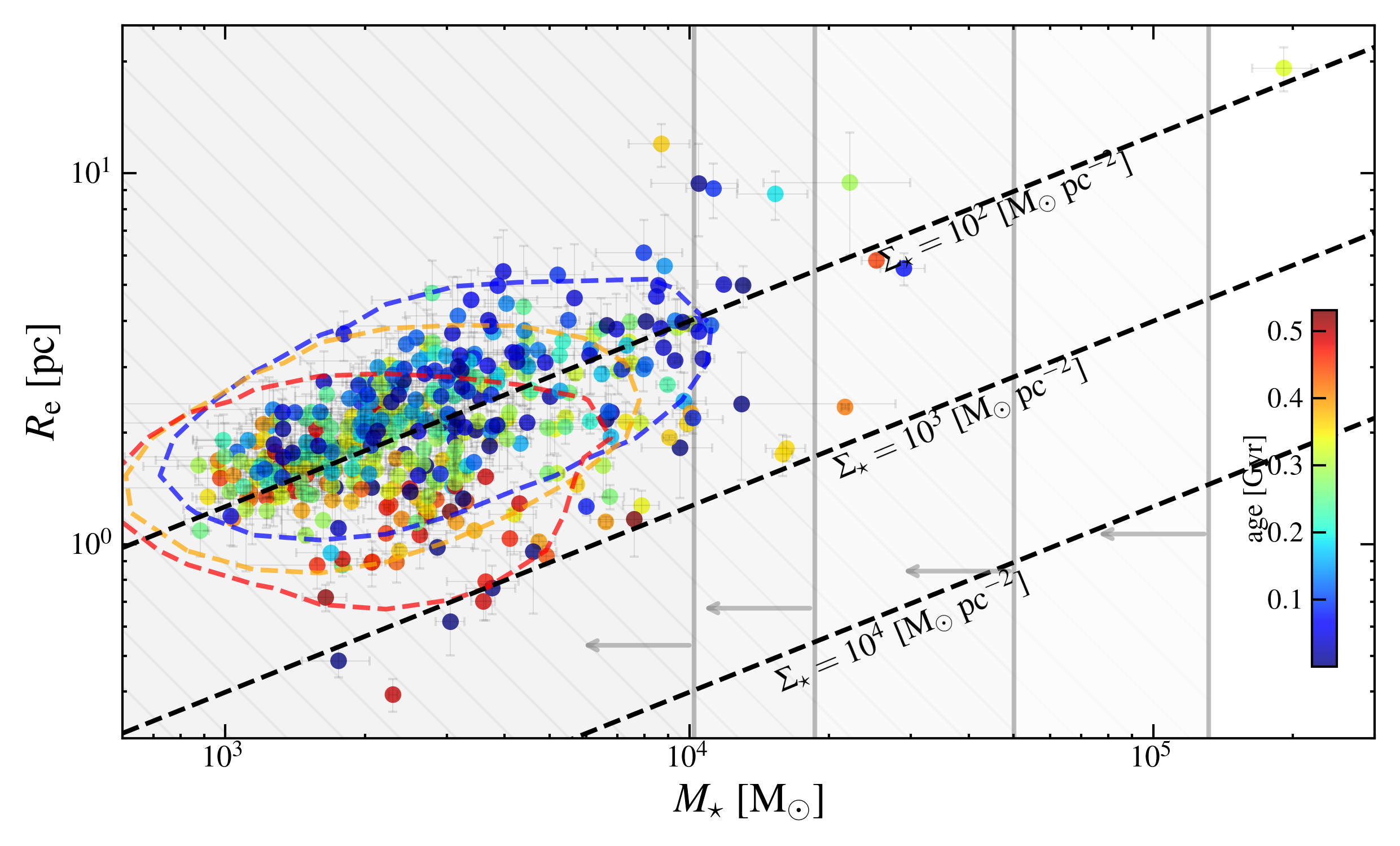}
    \caption{Size-mass relation at redshift 7, providing a more detailed view compared to Fig. 5. Clumps are color-coded by age, with blue indicating younger clumps and red representing older clumps. Clumps have been further grouped into age bins: clumps younger than $0.16\Gyr$, between $0.16\Gyr$ and $0.31\Gyr$ and older than $0.31\Gyr$, with each bin containing an equal number of objects. The isodensity contours for the three age distributions are also shown as dashed blue (young), yellow (intermediate-age) and red (old) lines. The grey vertical lines and bands represent the region of space where $\tevap\le13\Gyr$, with $\tevap$ as in equation~(\ref{for:evap}). Each line corresponds to a different assumption on $\RG$ in equation~(\ref{for:evap}), specifically, from left to right, $\RG=18\kpc$, $12\kpc$, $6\kpc$, $3\kpc$, together with $\VG=220\kms$ to mimic the tidal field of MW like galaxy (see text for details).}\label{fig:fig7_sizemassage}
\end{figure*}

\subsection{Clump dynamics and dissolution}
\label{subsec:evap}

In a simulation in which stars are treated as a collisionless component, there are fundamental limitations that prevent the accurate representation of certain physical phenomena. These limitations, by design, encompass collisional processes, which are essential when studying the long-term, dynamical evolution of dense star clusters and GCs. Being able to somewhat account for these phenomena is especially important in simulations like ours, where the dynamics of individual stars are resolved and high resolutions are achieved.

Two-body relaxation is a critical process in the dynamical evolution of star clusters (see e.g. \citealt{Spitzer1987}, \citealt{Heggie2003}). In densely packed environments, stars frequently engage in gravitational interactions, leading to a diffusion process that can cause stars to escape the cluster’s gravitational influence, a phenomenon leading to a decrease in the total mass of a GCs and in some cases to its eventual dissolution. Several factors can accelerate a cluster's evaporation and dissolution, such as the effects of the tidal truncation due to the external tidal field of the host galaxy, tidal shocks due the time variation of this tidal field, and stellar escape due to the cluster's expansion associated with mass loss due to stellar evolution and primordial gas expulsion \citep[see e.g.][for a review of various dynamical processes contributing to a cluster's evaporation]{Heggie2003}. A detailed investigation of the evolution of the clumps formed in our simulations is beyond the scope of this paper and will be presented in a future paper; here, we present only some general consideration and estimates on the dynamics and dissolution of these systems.

The results of a number of studies \citep[see e.g.][]{Chernoff1990,VesperiniHeggie1997,Vesperini1997,Baumgardt2003,GielesBaumgardt2008} have shown that mass loss due to two-body relaxation and the dissolution timescale depend mainly on the cluster's mass and the strength of the tidal field of the host galaxy. A cluster's dissolution may be further accelerated by early mass loss due to stellar evolution in particular for low-concentration clusters \citep[see e.g.][]{Chernoff1990,Fukushige1995}.

The complex and strong time variation in the tidal field in the high-redshift external environment like the one of our simulations can also significantly accelerate the mass loss and dissolution of clusters. An example of the dynamical effects associated with the high-redshift environment was presented in \cite{LiGnedin2019} who, on the basis of a semi-analytical calculation of the effects of these early time variation of the tidal field showed that most clusters either suffer a significant mass loss during these early evolutionary phases or undergo complete dissolution. In particular their analysis reveals that none of the clusters with masses smaller than $10^5\Msun$ survive until the present-day. Most of the clumps formed in our simulation have masses smaller than $\sim 2 \times 10^4\Msun$ and, although a detailed study of their dynamics will be addressed in a separate paper, we expect the vast majority of them to dissolve before $z=0$. Even without considering the effects of the time variation of the high-redshift tidal field, we can derive a very approximate estimate of the dissolution time of these clumps by using the expression of the dissolution timescale obtained by \cite{Baumgardt2003}
\begin{equation}\label{for:evap}
    \frac{\tevap}{\Myr} \equiv 1.03 \biggl[\frac{N}{\ln(0.02 N)}\biggr]^{0.82}\frac{\RG}{\kpc}\biggl(\frac{\VG}{220\kms}\biggr)^{-1}.
\end{equation}
In the above equation, $\RG$ represents the distance of the clump from the center of its host, $\VG$ is the circular velocity at that location, $N$ is the total number of stars within the clump. Calculating, just as an example, the dissolution time by adopting a mild tidal field equivalent to that of the Milky Way at $\RG=6\kpc$ (and $\VG=220\kms$), we find that all/vast majority of clumps dissolve before $z=0$: the fraction of disrupted clusters range from 100 per cent at $z=15.5$ and to 99.5 per cent at $z=6.14$\footnote{In the calculation $N$ is the ratio between the total clump mass and the average stellar mass, $0.6\Msun$ for a Kroupa IMF in the mass range $[0.1,40]\Msun$, the one samples by \citetalias{Calura2024}.}. Interestingly, at redshift 6.14, the sole clump with a dissolution timescale significantly exceeding $13\Gyr$ is the one already highlighted in Section \ref{subsec:cmf}, characterized by a mass of $10^6\Msun$ and distinct properties that set it apart from the rest of the population.

We strongly emphasize that this calculation provides just an approximate estimate of the timescale of clump dissolution but a more detailed modeling of the clump dynamics including additional processes, such as the tidal shocks associated with the time variation of the external tidal field, are to further accelerate the clump dissolution and confirm the large fraction of disrupted clusters before $z=0$ \citep[e.g.][]{LiGnedin2019}.

Finally, we conclude this section with a discussion of the possible role of clump mergers in the evolution of the clump population. Merging of smaller stellar systems into more massive clusters represents another critical process that not only affects the low-mass end but also may contribute to the enhancement over time of the high-mass end of the CMF. If mergers occur on timescales shorter than the dissolution time, they may further reduce the overall impact of dissolution on the CMF, despite its potential influence on low-mass populations of clusters. To quantify this aspect, we turn to Fig.~\ref{fig:fig7_sizemassage}, where we present a size-mass relation at $z=7$, similar to that shown in Fig.~\ref{fig:fig5_massfunction}, but now color-coded according to the clumps’ ages.  Blue indicates younger clumps, while red represents older clumps, with ages computed as described in Section~\ref{sec:struc}. The dashed isocontours reveal a clear trend with age: younger clumps tend to occupy regions of larger sizes and masses, while older clumps are found in areas with smaller sizes and masses. This age dependence indicates that the formation of more massive and larger structures is not primarily driven by merging of smaller clumps, which would populate the top-right part of the relation with older clumps. This is especially true in light of the typical star formation histories of the clumps. The middle panel of Fig.~\ref{fig:fig6_overallSFH} shows the star formation history of a selection of the clumps identified at $z=7$, with each color indicating a different clump. The SFHs of these clumps reveals sharp, short-lived episodes of star formation, with each burst lasting only a fraction of a $\Myr$. These SFHs generally display single peaks, rather than multiple bursts. To better highlight this aspect, the lower panels of Fig.~\ref{fig:fig6_overallSFH} present the distribution of $\log\sigmasf$ for all selected clumps at three representative redshifts. Here, $\sigmasf$ is defined as
\begin{equation}\label{for:sigsf}
    \sigmasf^2 = \frac{\int\SFH(t)(t-\tage)^2\dd t}{\int\SFH(t)\dd t},
\end{equation}
representing the dispersion in the SFH of a clump. At all redshifts, the distribution peaks around $\log\sigmasf/\Myr=-1$, with minimal spread, indicating that typical star formation episodes within clumps last approximately $0.1\Myr$ with little variability. However, each distribution shows a tail extending toward higher $\log\sigmasf$ values, with some reaching up to 1.5. These higher values do not reflect prolonged star formation episodes but rather arise from secondary peaks in the SFHs. These secondary peaks have systematically low star formation rates, at least two orders of magnitude weaker than the primary peak, but similar duration. As an example, in systems with masses of $10^4\Msun$ the secondary peak corresponds to the formation of only $\simeq100\Msun$, a negligible amount, comparable to that of a few massive stars. This small amount of mass is insufficient to be considered a substantial second star formation episode or accretion. Additionally, no clumps with masses exceeding $10^4\Msun$ exhibit multi-peaked SFH distributions, further underscoring that these secondary peaks represent minor contributions. This is a further clear indication that i) individual clumps rarely re-accrete gas to trigger subsequent star formation episodes and ii) that clumps are unlikely to result from mergers of multiple systems. 

We conclude that mergers are not the primary driver of clump growth. This is supported by the star formation histories, which show isolated, short bursts, suggesting that clumps rarely reaccrete gas or undergo frequent mergers, and how the size-mass relations are typically populated. Thus, dissolution remains an important process affecting the low-mass end of the CMF.

\section{Discussion}
\label{sec:disc}
In this Section, we discuss our results in the context of the implemented stellar feedback model (Section~\ref{subsec:feedback}), examining how it influences the formation and evolution of stellar clumps. We also compare our results with other simulation-based studies (Section~\ref{subsec:comparison}), providing a framework to assess the impact of different simulation set-ups on the properties of high-redshift stellar systems.

\subsection{The feedback model}
\label{subsec:feedback}

Among the simulations presented in \citetalias{Calura2024}, a star formation model with 100\% efficiency is necessary to create conditions suitable for producing massive and compact star clusters, underscoring the crucial impact of this parameter. Here, we provide a theoretical justification and describe the physical mechanism that facilitates the formation of these structures when $\epssfe=1$.

Recent studies (\citealt{Dekel2023}, but see also \citealt{Renzini2023,Renzini2025}) suggest that under appropiate conditions, stellar winds, radiative and supernova feedback may be insufficient to effectively heat or expel gas from star-forming regions, allowing star formation to proceed with remarkably high efficiency. This 'feedback-free' starburst (FFB) scenario requires a low-metallicity environment, typical at high redshift, and a gas density above a critical threshold ($3\times10^{3}\cm^{-3}$). Once this density is reached, the free-fall and cooling timescales become comparable to the delay in feedback effectiveness, enabling a rapid accumulation and conversion of gas into stars.

\begin{figure*}
    \centering
    \begin{minipage}{1\textwidth}
        \centering
        \includegraphics[width=1\hsize]{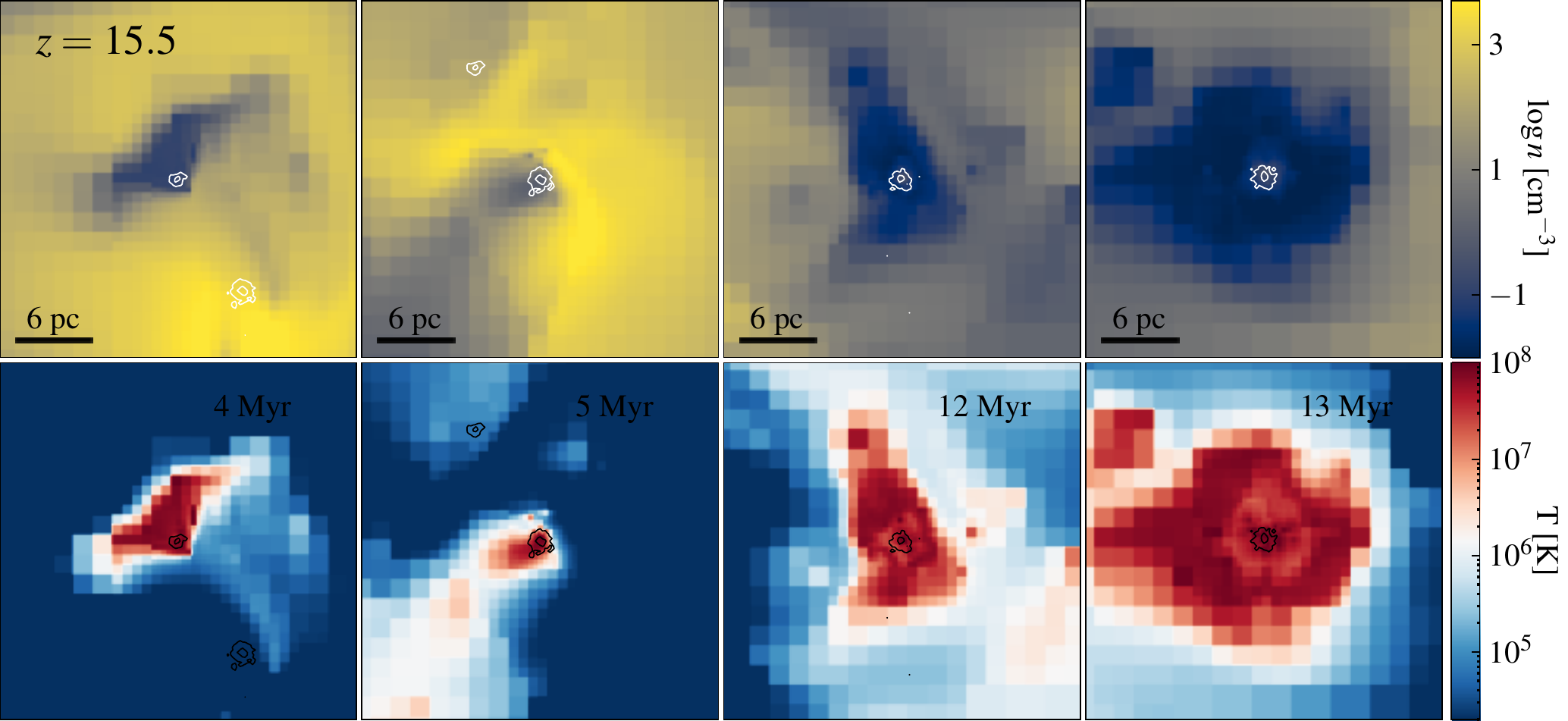}
        \vspace{0.05cm}
    \end{minipage}
    \begin{minipage}{1\textwidth}
        \centering
        \includegraphics[width=1\hsize]{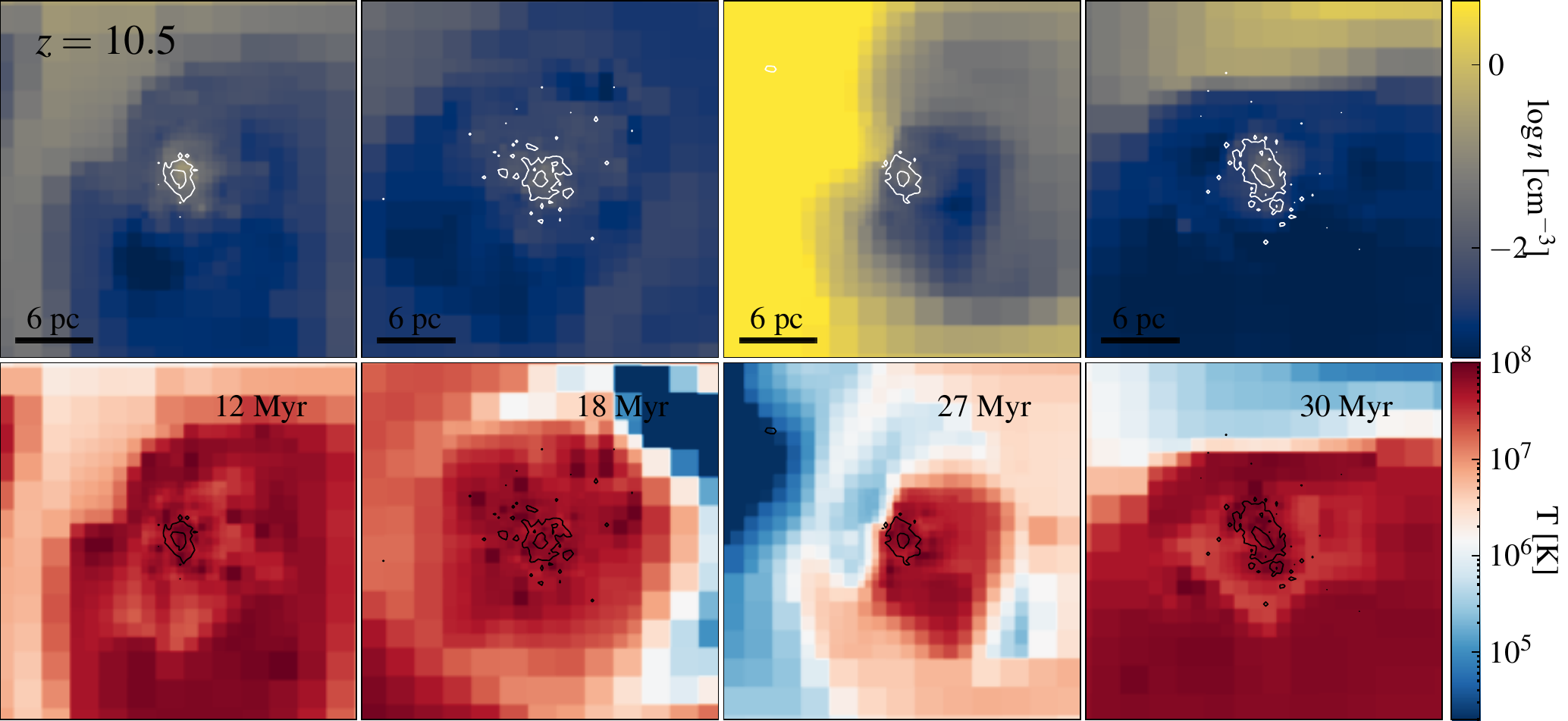}
    \end{minipage}
    \caption{
    Top set of panels: The first row presents a slice of gas density passing through the center of four selected clumps. Darker colors indicate regions of higher density, while lighter colors correspond to lower density areas. The second row displays the corresponding temperature maps. Superimposed on each map are curves of constant projected stellar density, shown as white lines (top) and black lines (bottom). All clumps in the first row are observed at redshift $z=15.5$. Bottom set of panels: same as the top panels, but for four clumps at $z=10.5$. In the top right part of each panel we have indicated the age of the clump.}
    \label{fig:fig8_gas}
\end{figure*}

Although differing in many respects, a FFB scenario and our star formation and feedback models share several similarities in the resulting phenomenology. In our simulations, star formation is triggered when gas cools below $2\times10^4\K$ and reaches a density threshold, which is higher than $10^4\cm^{-3}$ at the achieved resolution, exceeding the value required in the model by \cite{Dekel2023}. At these densities, the free-fall time decreases to $\tauff<0.5\Myr$, and the cooling time similarly shrinks to comparable values. Once stars form, since radiative feedback is not accounted for, feedback occurs through two channels only: stellar winds and supernovae. Supernovae become effective only after about $6\Myr$, corresponding to the lifetime of the most massive star ($40\Msun$) considered in the star formation model by \citetalias{Calura2024}. At this point, the differences in our model compared to the FFB framework are twofold: first, stellar winds inject energy and mass continuously; second, star formation efficiency remains fixed and does not depend on gas conditions.

For the FFB model to operate at this stage, a delayed onset of stellar winds would be necessary, which would automatically enhance the efficiency of star formation. In our case, setting $\epssfe=1$ accounts for this lack of self-consistency, enabling a rapid conversion of gas that proceeds until the injection of energy by winds is enough to heat adequately the remaining gas and bring the star formation episode to a halt. In our implementation, the thermal energy released by the winds increases rapidly as new stars are formed, and because this energy is substantial even at low metallicity, it effectively heats the surrounding gas on a short timescale.

Figure~\ref{fig:fig8_gas} shows density and temperature maps for selected clumps at redshifts $z=15$ (top rows) and $z=10.5$ (bottom rows) after the star formation episode in the clump is quenched. Each slice is taken on a plane passing through the center of the target clump, and young clumps were chosen to capture the feedback impact shortly after formation. Isodensity contours indicate projected stellar density along the corresponding line of sight. All clumps are embedded within large, high-temperature ($>10^7\K$) and low-density ($\log n[\cm^{-3}]<-1$) bubbles of gas recently heated by stellar winds and supernovae. In the two upper-left clumps, which are younger than $6\Myr$, feedback results exclusively from stellar winds, while in the other cases, both winds and supernovae contribute, as seen in the larger surrounding bubbles. This rapid, efficient star formation followed by intense heating drives the gas to conditions - high temperature and low density - unfavorable for further star formation. Consequently, the star formation history (SFH) becomes highly episodic, with brief, intense bursts \citep{Renzini2023}.

\subsection{Comparison with previous works}
\label{subsec:comparison}

This size-mass relation is a well-established trend, measured in observations \citep{Adamo2024} and, to some extent, reproduced by simulations. \cite{Ma2020} analyze high-resolution cosmological simulations at redshifts greater than 5 to study the formation of dense and bound stellar clumps at high redshift, possibly progenitor of present-day GCs. In their work the authors focused on the evolution of halos that reach virial masses of $10^{10}\Msun$, $10^{11}\Msun$ and $10^{12}\Msun$ by redshift 5 considering, also, different simulation resolutions. A key difference between our approaches is that \cite{Ma2020} do not model the formation of individual stars, but stellar particles, although of small mass, represent stellar populations. However, as in our case, they adopt a high star formation efficiency, even though the specific implementation of their star formation differs slightly from ours. \cite{Ma2020} find that all their simulated halos are able to form bound star clusters, which then follow a similar size-mass relation. However, their results show great sensitivity to resolution, which causes the minimum mass and the typical sizes of the clusters to be resolution dependent. Only their highest resolution simulation is comparable with ours and, in that case, for clumps with masses in the range of $\log\Mstar[\Msun]=3.5-4.5$, they report sizes between $2\pc$ and $100\pc$ (as shown in their Fig. 13), which overlap with the sizes we measure for clumps within the same mass range. However, their size distribution shows a significantly larger scatter compared to ours. Despite this, both their study and ours are in agreement on the CMF, reporting a power-law slope close to -2.

In another study, by means of hydrodynamical, cosmological simulations at sub-parsec resolution, \cite{Garcia2023} investigate the formation and evolution of star clusters in galaxies in the typical mass regime of dwarfs ($\Mvir\simeq2\times10^8$ at $z=8$). These simulations include radiative feedback, do not model individual stars as we do, and explore the effects of adopting two distinct star formation efficiencies: 0.35 (low) and 0.7 (high). The stellar clusters formed in their simulations span a lower mass range compared to ours, from approximately a few $10^2\Msun$ to $10^4\Msun$. As discussed by the authors, the CMF in their simulations changes significantly in response to repeated episodes of star formation, with the CMF flattening after each intense burst of new star formation. In their low-SFE run, \cite{Garcia2023} find a CMF slope similar to ours, but in their high-SFE case, the slope becomes shallower, around -0.6. This presents a key divergence from our results, as in our simulations, comparable to their high-SFE run, the CMF retains a steep slope of approximately $-2$. In their high-SFE run, the authors discussed how the bound clusters formed are systematically denser compared to the low-SFE case. This aligns with \citetalias{Calura2024}, where bound star cluster densities reach up to almost $10^4\Msun\pc^{-2}$ only in the high $\epssfe$ run, while they are much less dense and more diffuse in low $\epssfe$ simulations. As for the size-mass relation, \cite{Garcia2023} also observe a clear positive correlation. Additionally, their work finds a correlation between the size-mass relation and the ages of the clumps, indicating that younger clumps tend to populate the region of higher mass and larger size. This age correlation is consistent with our findings, further supporting the relationship between clump size, mass, and age.

\cite{Gutcke2024} presents high-resolution cosmological simulations aimed at exploring scenarios for GC formation. The simulations focus on dwarf galaxies with stellar masses, at $z=0$, ranging from $10^6\Msun$ to $10^7\Msun$, closely matching the properties of Local Group dwarf galaxies. The study reveals that clusters host ancient stellar populations and exhibit short, episodic star formation histories, occasionally marked by the presence of multiple stellar generations. Our size-mass relation is consistent with that of the author, showing a very similar correlation, although we note that \cite{Gutcke2024} forms clusters with a generally smaller average mass than ours. This may be attributed to the lower mass of the dark matter halo in their simulation. The author also finds very small fractions of dark matter relative to stellar mass in the stellar clusters formed, discussing how dark matter is lost through tidal stripping. While also the majority of our clusters live in environments free from dark matter, in our case, this absence is attributable to a formation without dark matter rather than a loss over time. Also, while the typical SFHs in \cite{Gutcke2024} and our clumps are characterized by short bursts of star formation, our clumps do not show the capability to reaccrete gas and lack multi-modal distributions, which is different from \cite{Gutcke2024}.

\section{Summary and conclusions}
\label{sec:concl}

In this study, we explore the formation of compact, high-redshift star-forming clumps using high-resolution cosmological zoom-in simulations. This simulation is part of the SIEGE project \citep{Calura2022}, a suite designed to investigate the conditions under which compact star clusters, possible progenitors of today's GCs, form. Specifically, this simulation is drawn from the larger suite described by \citetalias{Calura2024}, which examines the effects of feedback models, star formation efficiencies, and IMF variations on the structural and dynamical properties of young star clusters in high-redshift galaxies. The simulation targets a dwarf galaxy halo with a virial mass of $4\times10^9\Msun$ at $z=6.14$, achieving sub-parsec spatial resolution. This exceptional resolution, combined with individual star formation sampling from the IMF and an enforced 100\% star formation efficiency, enables us to investigate the conditions and mechanisms that govern star formation in the extreme environments characteristic of high-redshift proto-galaxies. In this work, we specifically investigate the properties of the stellar cluster population formed in our simulation, focusing on the evolution of their characteristics and scaling relations with redshift.

The high-redshift environment and the star formation and feedback model probed by our simulation naturally favor the formation of bound stellar clumps with masses spanning the range $10^3\Msun$ to $10^6\Msun$, with characteristic sizes on the order of 1-3 parsecs, and surface densities reaching up to almost $10^4\Msun\pc^{-2}$. The spatial resolution of the simulation ($\simeq0.2-0.5\pc$ across the redshift range) is comparable to the size of the smallest stellar clumps formed. As such, the clump sizes reported should be regarded as upper limits. While higher-resolution simulations could refine the structural characterization of clumps and potentially reveal more compact ones, such simulations are currently computationally unfeasible. Conversely, lower-resolution runs would artificially inflate clump sizes due to degraded gravitational accuracy, thus offering no meaningful convergence test. Remarkably, nearly all clumps are found to have minimal dark matter, with a ratio of dark-to-stellar mass within three effective radii below 1 in most cases. This finding suggests that such dense, compact structures can originate independently of dense dark matter halos and initially form as bound star clusters, rather than transient associations. Their eventual dissolution occurs on longer timescales, primarily due to interactions with the host galaxy. We reproduce the typical behavior of the CMF, which follows a power law of slope approximately -2. All clumps adhere to a mass-size relation that exhibits little evolution with redshift: smaller clumps tend to be less massive than larger clumps. The size mass relation is in agreement with results from other works. We find a clear trend in the size-mass relation with clumps age: younger clumps generally populate the higher mass and size regions. Clumps at higher redshift appear more compact, showing an evolution in surface density as we move to earlier cosmic times. The SFH of these clumps is consistently mono-modal, indicating well-defined ages and formation in single, rapid star formation episodes of typical duration of $\simeq0.1\Myr$. We observe no multi-modal SFHs, implying limited capacity for gas reaccretion and subsequent star formation bursts in these environments.

Our ability to resolve sub-parsec scales and incorporate feedback from individual stars is essential for capturing these small-scale processes that govern clump formation and evolution in high-redshift environments. This study provides a foundation for understanding the physical conditions under which GCs evolve, and it  demonstrates and highlights the critical role of star formation efficiency as a key parameter in formation models of high redshift clumps.

\begin{acknowledgements}

This paper is supported by the Italian Research Center on High Performance Computing Big Data and Quantum Computing (ICSC), project funded by European Union - NextGenerationEU - and National Recovery and Resilience Plan (NRRP) - Mission 4 Component 2 within the activities of Spoke 3 (Astrophysics and Cosmos Observations). The research activities described in this paper have been co-funded by the European Union - NextGeneration EU within PRIN 2022 project n.20229YBSAN - Globular clusters in cosmological simulations and in lensed fields: from their birth to the present epoch. We acknowledge support from the INAF Minigrant ‘Clumps at cosmological distance: revealing their formation, nature, and evolution' (Ob. Fu. 1.05.23.04.01). We acknowledge PRACE for awarding us access to Discoverer at Sofia Tech Park, Bulgaria. This research was supported in part by Lilly Endowment, Inc., through its support for the Indiana University Pervasive Technology Institute. We acknowledge EuroHPC JU for awarding the project IDs EHPC-REG-2021R0052 and EHPC-REG-2024R01-042 access to DISCOVERER at the Sofia Tech Park, Bulgaria. AL acknowledges support from PRIN MUR “2022935STW" funded by European Union-Next Generation EU, Missione 4 Componente 2
CUP C53D23000950006.
\end{acknowledgements}

\bibliographystyle{aa}
\bibliography{main}

\end{document}